\documentclass[a4paper]{amsart}
\usepackage{latexsym}
\usepackage{amsmath}
\usepackage{amssymb}
\usepackage{amsfonts}
\usepackage{esint}
\usepackage{graphicx}
\usepackage{verbatim}
\usepackage{color}
\definecolor{Blue}{rgb}{0.,0.,1.}
\definecolor{Red}{rgb}{1.,0.,0.}
\newcounter{smallarabics}
\newenvironment{arabicenumerate}
{\begin{list}{{\normalfont\textrm{(\arabic{smallarabics})}}}
  {\usecounter{smallarabics}\setlength{\itemindent}{0cm}
   \setlength{\leftmargin}{5ex}\setlength{\labelwidth}{4ex}
   \setlength{\topsep}{0.75\parsep}\setlength{\partopsep}{0ex}
   \setlength{\itemsep}{0ex}}}
{\end{list}}

\newcounter{smallroman}
\newenvironment{romanenumerate}
{\begin{list}{{\normalfont\textrm{(\roman{smallroman})}}}
  {\usecounter{smallroman}\setlength{\itemindent}{0cm}
   \setlength{\leftmargin}{5ex}\setlength{\labelwidth}{4ex}
   \setlength{\topsep}{0.75\parsep}\setlength{\partopsep}{0ex}
   \setlength{\itemsep}{0ex}}}
{\end{list}}

\newcommand{\ben}{\begin{arabicenumerate}}  
\newcommand{\een}{\end{arabicenumerate}}

\def\init{\setcounter{equation}{0}}

\newtheorem{theoreme}{Theorem }[section]
\newtheorem{proposition}[theoreme]{Proposition}
\newtheorem{lemma}[theoreme]{Lemma}
\newtheorem{definition}[theoreme]{Definition}
\newtheorem{corollary}[theoreme]{Corollary}
\newtheorem{remark}[theoreme]{Remark}
\newtheorem{example}[theoreme]{Example}
\newcommand{\beq}{\begin{equation}}
\newcommand{\eeq}{\end{equation}}

\newcommand{\bex}{\begin{example}}
\newcommand{\eex}{\end{example}}
\def\bel{\begin{lemma}}
\def\eel{\end{lemma}}
\def\bet{\begin{theoreme}}
\def\eet{\end{theoreme}}
\def\bed{\begin{definition}}
\def\eed{\end{definition}}
\def\ber{\begin{remark}}
\def\eer{\end{remark}}

\def\rr{{\mathbb R}}

\def\cc{{\mathbb C}}
\def\nn{{\mathbb N}}

\def\part{{\rm par}}

\def\Re{{\rm Re}}

\def\bar{\overline}

\def\cinf{C^\infty}
\def\c0inf{C_0^\infty}

\def\proof{
\noindent{\bf Proof.}\ \ }

\def\ch{{\mathfrak h}}

\def\cZ{{\mathcal Z}}
\def\cY{{\mathcal Y}}

\def\cS{{\mathcal S}}

\def\cD{{\mathcal D}}

\def\cM{{\mathcal M}}
\def\cN{{\mathcal N}}

\def\i{{\rm i}}
\def\ad{{\rm ad}}

\def\Dom{{\rm Dom}}

\def\T{{\rm T}}

\def\Ker{{\rm Ker}}

\def\qed{$\Box$\medskip}

\def \p{ \partial}

\def\12{\frac{1}{2}}
\def\14{\frac{1}{4}}

\def\supp{{\rm supp}}

\def\ad{{\rm ad}}
\def\e{{\rm e}}

\def\d{{\rm d}}

\def\bbbone{{\mathchoice {\rm 1\mskip-4mu l} {\rm 1\mskip-4mu l}
{\rm 1\mskip-4.5mu l} {\rm 1\mskip-5mu l}}}
\def\one{\bbbone}
\def\cH{{\mathcal H}}

\def\ii{{\rm j}}
\def\jj{ {\rm j}_1}

\def\w{{\rm w}}

\def\Ker{{\rm Ker}}

\def\coinf{C_0^\infty}

\def\cG{{\mathcal G}}
\def\cX{{\mathcal X}}

\def \p{ \partial}

\def\12{\frac{1}{2}}

\def\supp{{\rm supp}}

\def\ad{{\rm ad}}
\def\e{{\rm e}}

\def\d{{\rm d}}

\def\cH{{\mathcal H}}

\def\bep{\begin{proposition}}
\def\eep{\end{proposition}}

\def\w{{\rm w}}

\def\Opw{{\rm Op}^{\rm w}}

\def\Op{{\rm Op}}

\newcommand{\mat}[4]{\left(\begin{array}{cc}#1 &#2  \\ #3 &#4 \end{array}\right)}

\def\CARal{{\rm C\hskip 0.25 em \hbox{\raise 1.72 ex 
\hbox{$\scriptscriptstyle\rm al$}\kern -0.57 em A}R}}

\def\t{{\scriptscriptstyle\#}}
\def\otimesal{\mathop{\hbox{\raise 1.5 ex
  \hbox{$\scriptscriptstyle\rm al$}
\kern -0.92 em \hbox{$\otimes$}}}}
\def\oplusal{\mathop{\hbox{\raise 1.5 ex
  \hbox{$\scriptscriptstyle\rm al$}
\kern -0.92 em \hbox{$\oplus$}}}}
\def\Gammal{\hbox{\raise 1.68 ex 
\hbox{$\scriptscriptstyle\rm al$}\kern -0.50 em $\Gamma$}}
\def\Bal{\hbox{\raise 1.68 ex 
\hbox{$\scriptscriptstyle\rm  al$}\kern -0.50 em $B$}}
\def\CARal{{\rm C\hskip 0.25 em \hbox{\raise 1.72 ex 
\hbox{$\scriptscriptstyle\rm al$}\kern -0.57 em A}R}}
\def\t{{\scriptscriptstyle\#}}

\def\cE{{\mathcal E}}

\def\rx{{\rm x}}
\def\rA{{\rm A}}\def
\41{\frac{1}{4}}
\newcommand\Texp{{\rm  Texp}}
\def\WF{{\rm WF}}
\newcommand{\co}[1]{T^{*}\! #1}
\newcommand{\coo}[1]{T^{*}\! #1\backslash Z}
\newcommand{\cooo}[1]{T^{*} #1\backslash \{0\}}

\newcommand{\shalf}{{\textstyle \frac{1}{2}}}
\newcommand{\half}{{\frac{1}{2}}}
\newcommand{\kg}{{\rm (KG)}}

\newcommand{\bra}{\langle} 
\newcommand{\ket}{\rangle}

\def\shit{(\rr^{d})\otimes \cc^{2}}

\begin{document}
\title[Construction of Hadamard states]{Construction of Hadamard states by pseudo-differential calculus}
\author{C. G\'erard}
\address{D\'epartement de Math\'ematiques, Universit\'e de Paris XI, 91405 Orsay Cedex France}
\email{christian.gerard@math.u-psud.fr}
\author{M. Wrochna}
\address{Mathematisches Institut, Universit\"at G\"ottingen, Bunsenstr. 3-5, D - 37073 G\"ottingen, Germany}
\email{wrochna@uni-math.gwdg.de}
\begin{abstract}
 We give a new construction based on pseudo-differential calculus of quasi-free Hadamard states for Klein-Gordon equations on a class of space-times whose metric is well-behaved at spatial infinity. In particular on this class of space-times,  we construct all pure Hadamard states whose two-point function (expressed in terms of Cauchy data on a Cauchy surface) is a matrix of pseudo-differential operators.  We also study their covariance under symplectic transformations. 
 
 As an aside, we give a new construction of Hadamard states on arbitrary globally hyperbolic space-times which is an alternative to the classical construction by Fulling, Narcowich and Wald.
\end{abstract}
\maketitle

\section{Introduction}\label{sec:introduction}
\subsection{Hadamard states}
Hadamard states are nowadays widely accepted as possible physical states of the non-interacting quantum field theory on a curved space-time. One of the main reason is their applicability to renormalization of the stress-energy tensor, a necessary step in the formulation of semi-classical Einstein equations. Moreover, the Hadamard condition plays an essential role in the perturbative construction of interacting quantum field theory \cite{BF00}. Other related concepts making use of Hadamard states include local thermal equilibrium \cite{schlemmer} and quantum energy inequalities \cite{fewster}. 

Since the work of Radzikowski \cite{radzikowski}, the Hadamard condition (renamed \emph{microlocal spectrum condition}), is formulated as a requirement for the wave front set of the associated two-point function $\Lambda$, which is necessarily a bi-solution of the free equations of motion. It is therefore natural to try to construct such states using the standard apparatus of microlocal analysis, based on pseudo-differential calculus. Although a construction is already known for space-times with compact Cauchy surface \cite{junker}, it does certainly not cover many cases of physical interest and lacks the capability to produce many states on a fixed space-time with distinct properties.

 In this paper we address these questions for  a class of space-times whose metric components are suitably well-behaved at spatial infinity, allowing also for external potentials. In this case  it is possible to obtain rather complete and transparent results.
 Namely we construct a large class of quasi-free Hadamard states, whose two-point functions, expressed in terms of Cauchy data on a fixed Cauchy surface, are matrices of pseudo-differential operators.
 In particular we can construct \emph{all } such \emph{pure} Hadamard states and study their covariance under symplectic transformations.
 
 As an additional result we give a new construction of Hadamard states on arbitrary globally hyperbolic space-times which is an alternative to the classical construction by Fulling, Narcowich and Wald \cite{FNW}. Our method turns out not to produce pure states in general, it allows however to keep track of their local properties.
 
\subsection{Methods}
Our analysis is set on three levels:
\begin{enumerate}\setlength{\leftmargin}{5ex}\setlength{\itemindent}{0cm}
\item\label{it:i1} Our starting point are normally hyperbolic operators on $\rr\times\rr^d$ of the form
\beq\label{asto}
\p_{t}^{2}+ a(t, \rx, D_{\rx})=\p_t^2 -\p_{\rx^{j}}a^{jk}\p_{\rx^{k}}+  b^{j}\p_{\rx^{j}}- \p_{\rx^{j}}\overline{b}^{j}+ m,
\eeq
where 
\beq\label{asto-hyp}
\begin{array}{l}
a^{jk}, b^{j}, m \in C^{\infty}(\rr, C^{\infty}_{\rm bd}(\rr^{d})), \   m(x)\in \rr,\\[2mm]
[a^{jk}](x)\geq  c(t)\one\hbox{ \  uniformly on }\rr^{1+d}, \ c(t)>0. 
\end{array}.
\eeq
We refer to this case as the \emph{model Klein--Gordon equation} and give a construction of the associated parametrix  for  the Cauchy problem, in such way that the propagation of positive-frequency and negative-frequency singularities is under control. This allows us to reformulate the microlocal spectrum condition in terms of Cauchy data. We show how to construct many non-necessarily pure Hadamard states and then characterize pure ones. We also describe classes of symplectic transformations which preserve the microlocal spectrum condition.

\item\label{it:i2} The above results are easily extended to operators of the form $f (\p_{t}^{2}+ a(t) )g$, where $f$ and $g$ are smooth densities. This way, we show that the problem of constructing Hadamard states is reduced to the model case above if $M=\rr\times\rr^d$, the metric is given by
\beq\label{asti}
g= -c (x)dt^{2}+ h_{jk}(x)d\rx^{j}d\rx^{k},
\eeq
and the Klein-Gordon operator is of the form 
\[\begin{array}{rl}
P(x, D_{x})=&\ c^{-\12}|h|^{-\12}(\p_{t}+ \i V) c^{-\12}|h|^{\12}(\p_{t}+\i V)\\[2mm]
&- c^{-\12}|h|^{-\12}(\p_{j}+\i A_{j})c^{\12}| h|^{\12}h^{jk}(\p_{k}+\i A_{k})+ \rho,
\end{array}
\]
where 
$A_{\mu}(x)= (V(x), \rA_{j}(x))$, $|h|= \det[h_{jk}]$, $ [h^{jk}]= [h_{jk}]^{-1}$ and the following hypotheses are assumed:
\beq\label{asti-hyp}
 \begin{array}{rl}
&\forall \ I\subset \rr\hbox{ compact interval }\exists\  C>0\hbox{ such that }\\[2mm]
&C\leq c(x), \ C\one\leq [h_{jk}(x)],\hbox{uniformly for }x\in I\times \rr^{d},\\[2mm]
&  h_{jk}(x), \ c(x), \ \rho(x), A_{\mu}(x)\in C^{\infty}(\rr, C^{\infty}_{\rm bd}(\rr^{d})).
\end{array}
\eeq
\item\label{it:i3} For arbitrary space-times (and external potentials), using a suitable partition of unity, we explain how to glue together two-point functions of Hadamard states on smaller regions of the space-time into a globally-defined one. Using the results obtained for the special case above, this yields a new construction of Hadamard states on arbitrary globally-hyperbolic space-times.
\end{enumerate}

Let us mention that beside the construction for space-times with compact Cauchy surface due to Junker, a general existence result for Hadamard states is known \cite{FNW}, as well as a collection of various examples on specific classes of space-times. However, the existence argument of Fulling, Narcowich and Wald has the disadvantage of being highly non-explicit and requires non-local information on the space-time as an input. Those drawbacks are to a large extent avoided in our approach, as explained in Section \ref{sec7}. As for the known examples of Hadamard states for the Klein-Gordon equation, these include:
\begin{romanenumerate}
\item passive states for stationary space-times (this includes ground- and KMS states) \cite{passivity},
\item states constructed in \cite{DMPas} for a subclass of asymptotically flat vacuum space-times at null infinity (see \cite{moretti} for the proof),
\item states constructed in \cite{DMPcosm} for a class of cosmological space-times (this includes the Bunch-Davies state on de-Sitter space-time),
\item so-called states of low energy for FLRW space-times \cite{olbermann},
\item the so-called Unruh state \cite{unruh},
\item ground states and over-critical states for static potentials on Minkowski space-time considered in \cite{wrochna}.
\end{romanenumerate}
A short inquiry shows that the sets of assumptions (\ref{asto-hyp}) and (\ref{asti-hyp}) studied by us in greater detail are only partially covered by the examples above.

\subsection{Plan of the paper}

The paper is organized as follows.

In Section \ref{sec0} we recall basic facts on bosonic quasi-free states. A special emphasis is put on explaining the relation between the neutral and the (less often discussed) {\em charged} case. In our problem, the use of microlocal analysis makes much more natural to work with complex quantities. In order to cover both cases it is sufficient to consider gauge-invariant charged quasi-free states. 

Section \ref{sec22} contains basic definitions and facts on the Klein-Gordon equation, wave front sets and Hadamard states.

 In Section \ref{sec1} we recall mostly well-known results on pseudo-differential calculus needed later on. This includes theorems on the pseudo-differential property of functions of pseudo-differential  operators and several results related to Egorov's theorem. In Section \ref{sec2} we specify our assumptions for the space-times (\ref{asti}) and explain the reduction to the model Klein-Gordon equation (\ref{asto}).

Section \ref{sec3} contains the key technical ingredient used to construct Hadamard states in Section \ref{sec4},  namely a sufficiently explicit construction of a {\em parametrix} for the Cauchy problem.  Let us explain in  more details  this construction and its use.

As is well-known, the appropriate phase-space for the quantization of the Klein-Gordon equation is the space of its smooth {\em space-compact solutions}, which can be identified with  the image of $\cD(M)$  under the Pauli-Jordan commutator function $E$. 
Alternatively one can fix  a Cauchy surface $\Sigma\subset M$ and identify a space-compact solution $\phi$ with its Cauchy data $\rho\phi= (\rho_{0}\phi, \rho_{1}\phi)\in \cD(\Sigma)\oplus \cD(\Sigma)$.

The Hadamard condition on a state $\omega$ is formulated in terms of its  two-point function $\Lambda$, which is an element of $\cD'(M\times M)$.  For practical purposes it is more convenient to fix a Cauchy surface $\Sigma\subset M$ and to consider instead  the two-point function $\lambda$ of $\omega$ in terms of Cauchy data. Both two-point functions are related by:
\beq\label{covcov}
 \Lambda= (\rho\circ E)^{*} \circ \lambda \circ (\rho\circ E).
\eeq
Since the Hadamard condition singles out the {\em positive energy} component  $\cN_{+}$ of the characteristic manifold $\cN= \{(x, \xi)\in T^{*}M: \xi_{\nu}g^{\mu\nu}(x)\xi_{\mu}=0\}$, we see from (\ref{covcov}) that the essential step is to formulate conditions on the Cauchy data $\rho \phi$ which ensure that a (distributional or finite energy) solution $\phi$ will have its wave front set contained in $\cN_{+}$.
The natural way to do this is to construct a {\em parametrix} for the Cauchy problem on $\Sigma$, i.e. a solution  modulo smooth errors of the Cauchy problem with data $(f_{0}, f_{1})$ compactly supported distributions.  

Such a construction is well-known in microlocal analysis and is usually done with  {\em Fourier integral operators}.   To obtain more transparent results later on,  we use instead  a more abstract construction, relying on Hilbert space methods, which leads to a smaller number of arbitrary choices. In our case the arbitrariness is described by the choice of a pseudo-differential operator $r$ acting on $\Sigma$ (see Def. \ref{defo}).  A concrete consequence of our parametrix construction is as follows:  let us consider for definiteness the space ${\rm Sol}_{E}(P)$ of {\em finite energy solutions} of the Klein-Gordon equation(see Subsect. \ref{sec3.3}). Then for each choice of $r$ this space splits into the  direct sum of two spaces ${\rm Sol}_{E}^{\pm}(P, r)$, whose elements have wavefront sets in $\cN_{\pm}$ and which are {\em symplectically orthogonal}.  We will use this fact through an appropriate reparametrization of Cauchy data in which the symplectic form still has a very simple form (see Lemma \ref{4.1}).

Section \ref{sec4} contains the key results of the paper. Theorem \ref{4.2} characterizes Hadamard states for the model Klein--Gordon equation in terms of their symplectically smeared two-point function $\lambda$. This allows us to construct a large class of Hadamard states in Subsect. \ref{4.3}. In Subsect. \ref{sec4.4} we describe classes of symplectic transformations which preserve the microlocal spectrum condition. Stronger results are derived for pure quasi-free states in Subsect. \ref{sec4.5}, in particular {\em all} pure Hadamard states with pseudo-differential two-point functions are characterized.

Among these pure states, there exists  a `canonical' Hadamard state  with two-point function $\lambda(r)$, which is distinguished modulo the choice of the pseudo-differential operator $r$ appearing in the construction of the parametrix. We explicitely find the symplectic transformation relating two canonical two-point functions  $\lambda(r_{1})$, $\lambda(r_{2})$ for two different choices of $r$. In this way the remaining ambiguities in the construction of $\lambda(r)$ are completely understood.

In Subsect. \ref{for-record} we briefly discuss the static case and show how the ground state and KMS states fit in our construction.

In Section \ref{sec7} we present our  alternative construction of Hadamard states on an arbitrary globally hyperbolic space-time. Instead of the classical deformation argument  of \cite{FNW}, we fix a Cauchy surface $\Sigma$ and use the fact that the symplectic form expressed in terms of Cauchy data is a local operator.

We include some remarks on the case of a compact Cauchy surface and compare our results with the construction from \cite{junker} in Subsect. \ref{ssec:junker}. Various proofs are collected in Appendix \ref{sec-app1}.

\section{Bosonic quasi-free states}\label{sec0}\init

In this section we recall well-known facts about bosonic quasi-free states, following \cite {DG}.  We choose to  work with complex symplectic spaces and gauge invariant states  (or equivalently charged fields), possibly after complexification of a  real situation. In this respect the structures obtained by complexifying a complex symplectic space deserves some attention, see  Subsect. \ref{complix}.

\subsection{Notation}\label{notation}
\def\dito{\!\cdot\!}
If $\cX$ is a real or complex vector space we denote by $\cX^{\t}$ its dual. Bilinear forms on $\cX$ are identified with elements of $L(\cX, \cX^{\t})$, which leads to the notation $x_{1}\!\cdot \!bx_{2}$ for $b\in L(\cX, \cX^{\t})$, $x_{1}, x_{2}\in \cX$. The space of symmetric (resp. anti-symmetric) bilinear forms on $\cX$ is denoted by $L_{\rm s}(\cX, \cX^{\t})$ (resp. $L_{\rm a}(\cX, \cX^{\t})$). 

If $\sigma\in L_{\rm s}(\cX, \cX^{\t})$, we denote by $O(\cX, \sigma)$ the linear (pseudo-)orthogonal group on $\cX$. Similarly if
 $\sigma\in L_{\rm a}(\cX,\cX^{\t})$ is non-degenerate, i.e. $(\cX, \sigma)$ is a symplectic space, we denote by $Sp(\cX, \sigma)$ the linear symplectic group on $\cX$.

If $\cX$ is a complex vector space,  we denote by $\cX_{\rr}$ its {\em real form}, i.e. $\cX$ considered as a real vector space. 
We denote by $\overline{\cX}$ a {\em conjugate vector space} to $\cX$, i.e. a complex vector space $\overline{\cX}$ with an anti-linear isomorphism
$\cX\ni x\mapsto \overline{x}\in \overline{\cX}$. The {\em canonical conjugate vector space} to $\cX$ is simply  the real vector space $\cX_{\rr}$ equipped with the complex structure $-\i$, if $\i$ is the complex structure of $\cX$. In this case the map $x\to \overline{x}$ is the identity.
If $a\in L(\cX_{1}, \cX_{2})$, we denote by $\overline{a}\in L(\overline{\cX}_{1}, \overline{\cX}_{2})$ the linear map defined by:
\beq\label{toti}
\overline{a}\overline{x}_{1}:= \overline{ax_{1}}, \ \overline{x}_{1}\in \overline{\cX}_{1}.
\eeq

We denote by $\cX^{*}$ the {\em anti-dual} of $\cX$, i.e. the space of anti-linear forms on $\cX$. 
Clearly $\cX^{*}$ can be identified with $\overline{\cX^{\t}}\sim \overline{\cX}^{\t}$.

Sesquilinear forms on $\cX$ are identified  with elements of $L(\cX, \cX^{*})$, and we use the notation $(x_{1}|b x_{2})$ or sometimes $\overline{x_{1}} \dito b x_{2}$ for $b\in L(\cX, \cX^{*})$, $x_{1}, x_{2}\in \cX$.

The space of hermitian (resp. anti-hermitian) sesquilinear forms on $\cX$ is denoted by $L_{\rm s}(\cX, \cX^{*})$ (resp. $L_{\rm a}(\cX, \cX^{*})$).

If $q\in L_{\rm s}(\cX, \cX^{*})$ is non-degenerate, i.e. $(\cX, q)$ is a pseudo-unitary space, we denote by $U(\cX, q)$ the linear pseudo-unitary group on $\cX$.

If $b$ is a bilinear form on  the real vector space $\cX$, its canonical sesquilinear extension to $\cc\cX$ is by definition the sesquilinear form $b_{\cc}$ on $\cc\cX$ given by
\[
(w_1|b_{\cc}w_2):=x_1 \dito b x_2+y_1  \dito b y_2+\i x_1 \dito  b y_2 - \i y_1  \dito b x_2, \quad w_i=x_i+\i y_i
\]
for $x_i,y_i\in\cX$, $i=1,2$. This extension maps (anti-)symmetric forms on $\cX$ onto (anti-)hermitian forms on $\cc\cX$.

Conversely if $\cX$ is a complex vector space and $\cX_{\rr}$ is its real form, i.e. $\cX$ considered as a real vector space,  then for $b\in L_{\rm s/a}(\cX, \cX^{*})$ the form ${\rm Re }b$ belongs to $L_{\rm s/a}(\cX_{\rr}, \cX_{\rr}^{\t})$.

\subsection{Bosonic quasi-free states:  neutral case}\label{sec0.1}
Let $(\cX, \sigma)$ be a real symplectic space, i.e. a pair consisting of a real vector space $\cX$ and a non-degenerate anti-symmetric form $\sigma\in L_{\rm a}(\cX, \cX^{\t})$.

We denote $\mathcal{A}(\cX,\sigma)$ the Weyl CCR $C^*$-algebra of $(\cX, \sigma)$, formally generated by elements of the form $W(y)$ for $y\in\cX$,  with:
\[ 
W(y)^* = W(-y), \ \ \  W(x)W(y) = \e^{-\i ( x \dito  \sigma y)/2} W(x+y), \ \ \ x,y\in\cX. 
\]

\begin{definition}A state $\omega$ on $\mathcal{A}(\cX,\sigma)$ is called a (bosonic, neutral) quasi-free state if there is a symmetric form $\eta$ (called  the {\em covariance} of $\omega$) on $\cX$ such that
\[
\omega(W(x))=\e^{-\half x \dito \eta x}, \ \ \ \ \ x\in\cX.
\]
\end{definition}
A quasi-free state $\omega$ on $\mathcal{A}(\cX, \sigma)$ is regular, i.e. the field operators $\phi(x)$ are well-defined as selfadjoint operators in the GNS representation of $\omega$ with:
\[
[\phi(x_{1}), \phi(x_{2})]= \i x_{1} \dito  \sigma x_{2}\one,\hbox{ as quadratic forms on }\Dom \phi(x_{1})\cap \Dom \phi(x_{2}),
\]
and:
\begin{equation}
\label{e0.000}
\omega(\phi(x_{1})\phi(x_{2}))= x_{1}  \dito \eta x_{2}+ \frac{\i}{2} x_{1} \dito  \sigma x_{2}, \quad x_{1}, x_{2}\in \cX.
\end{equation} 
It is convenient to introduce the sesquilinear hermitian form
\[
q:= \i \sigma_{\cc},
\]
usually called the {\em charge} and
\[
\lambda_{\pm}:= \eta_{\cc}\pm  \12 q \in L_{\rm s}(\cc\cX, \cc\cX^{*}).
\]
The following results are well-known (see e.g. \cite[Chap. 17]{DG}).
\begin{proposition}\label{prop:defomega} Let $\eta\in L_{\rm s}(\cX, \cX^{\t})$. Then  the following are equivalent:
\ben
\item $\eta$ is the covariance of a quasi-free state on $\mathcal{A}(\cX,\sigma)$,
\item  
$x   \dito \eta x\geq 0$, $|x_{1}  \dito \sigma x_{2}|\leq 2 (x_{1} \dito \eta x_{1})^{\12}(x_{2} \eta x_{2})^{\12}$, $x_{1},x_{2}\in\cX$,
\item $\lambda_{\pm}\geq 0$ on $\cc\cX$.
\een
\end{proposition}

\begin{proposition}\label{ista}
Let $\eta\in L_{\rm s}(\cX, \cX^{\t})$.  Then  the following are equivalent:
\ben
\item $\eta$ is the covariance of a {\em pure} quasi-free state on $\mathcal{A}(\cX,\sigma)$,
\item   $(2\eta, \sigma)$ is {\em K\"{a}hler}, i.e.  there exists an anti-involution $\jj\in Sp(\cX, \sigma)$ such that  $2\eta= \sigma \jj$.
\een
\end{proposition}
\begin{proposition}\label{istaa}
 Let $\eta_{1}$, $\eta_{2}$ be covariances of two pure quasi-free states on $\mathcal{A}(\cX,\sigma)$. Then there exists $r\in Sp(\cX, \sigma)$ such that $\eta_{2}= r^{\t}\eta_{1}r$.
\end{proposition}

\subsection{ Bosonic quasi-free states: charged case}\label{sec0.2}

Let us now consider the case of a complex symplectic space $(\cY, \sigma)$, i.e.  a pair consisting of a complex vector space $\cY$ and a non-degenerate anti-hermitian form $\sigma\in L_{\rm a}(\cY, \cY^{*})$. 
The complex structure on $ \cY$ will be denoted by $ \ii$, to distinguish it from the complex number $ \i\in \cc$.

 As before we introduce  the charge $q:= \i \sigma$ which is  hermitian.

Note that  $(\cY_{\rr}, {\rm Re}\sigma)$ is a real symplectic space called the {\em real form } of $(\cY, \sigma)$, with $\ii\in Sp(\cY_{\rr}, {\rm Re}\sigma)$.

Conversely if $(\cX, \sigma)$ is a real symplectic space equipped with an anti-involution  $\ii\in Sp(\cX, \sigma)$, then denoting by $\cY$ the space $\cX$ equipped with the complex structure $\ii$ and setting $(x_{1}| \hat{\sigma}x_{2}):= x_{1} \dito \sigma x_{2}- \i  x_{1}  \dito \sigma \ii x_{2}$, the space $(\cY, \hat{\sigma})$ is a complex symplectic space whose real form is $(\cX, \sigma)$.
 
For coherence of notation we will denote the Weyl CCR algebra $\mathcal{A}(\cY_{\rr}, \Re \sigma)$ by $\mathcal{A}(\cY, \sigma)$.

Let us now consider a quasi-free state $\omega$ on $\mathcal{A}(\cY, q)$, as in Subsect. \ref{sec0.1}. The state $\omega$ is called \emph{gauge-invariant} if
\[
\omega(W(y))=\omega(W(\e^{\ii\theta}y)), \ \ 0\leq\theta<2\pi, \ \ y\in\cY.
\]

If the state $\omega$ is not gauge-invariant, the complex structure $\ii$ plays no role and one can forget it. One is then reduced to the situation of Subsect. \ref{sec0.1}.

If $\eta$ is the covariance of $\omega$ then $\omega$ is gauge-invariant iff $\ii\in U(\cY_{\rr}, \eta)$. One can then uniquely associate to $\eta$ a $\ii-$sesquilinear hermitian form $\hat{\eta}$ defined by
\begin{equation}
\label{tati}
(y_{1}| \hat{\eta}y_{2}):= y_{1} \dito \eta y_{2}-\i y_{1} \dito  \eta\ii y_{2}, \ y_{1}, y_{2}\in \cY.
\end{equation}

Let now $\phi(y)$ for $y\in \cY$ be  the  selfadjoint fields in the GNS representation of $\omega$. One can  introduce  the {\em charged fields}:
\[
\psi(y):=\frac{1}{\sqrt{2}}(\phi(y)+ \i \phi(\ii y)), \ \psi^{*}(y):=\frac{1}{\sqrt{2}}(\phi(y)- \i \phi(\ii y)), \ y\in \cY.
\]
The map $\cY\ni y\mapsto \psi^{*}(y)$ (resp. $\cY\ni y\mapsto \psi(y)$) is $\cc-$linear (resp. $\cc-$anti-linear). The commutation relations take the form:
 \[
[\psi(y_{1}), \psi(y_{2})]= [\psi^{*}(y_{1}), \psi^{*}(y_{2})]=0,  \ [\psi(y_{1}), \psi^{*}(y_{2})]=  (y_{1}| qy_{2})\one, \ y_{1}, y_{2}\in \cY.
\]
If $\omega$ is a gauge-invariant quasi-free state on $\mathcal{A}(\cY, q)$, then:
\[
\omega(\psi(y_{1})\psi(y_{2}))= \omega(\psi^{*}(y_{1})\psi^{*}(y_{2}))=0, \ y_{1}, y_{2}\in \cY,
\]
and we set:
\beq\label{e0.tutu}
\begin{array}{rl}
\omega(\psi(y_{1})\psi^{*}(y_{2}))=: &(y_{1}| \lambda_{+} y_{2}),\\[2mm]
\omega(\psi^{*}(y_{2})\psi(y_{1}))=: &(y_{1}| \lambda_{-} y_{2}), \ y_{1}, y_{2}\in \cY.
\end{array}
\eeq
Clearly $\lambda_{+}-\lambda_{-}=q$.  

We will call $\lambda_{\pm}\in  L_{\rm s}(\cY, \cY^{*})$ the 
{\em complex covariances } of the gauge invariant quasi-free state $\omega$.

 Introducing the selfadjoint fields $\phi(y)$ we obtain that
\[
\omega(\phi(y_{1})\phi(y_{2}))= {\rm Re}(y_{1}| (\lambda- \12 q)y_{2})+ \frac{\i}{2}{\rm Re}(y_{1}| \sigma y_{2}).
\]
Therefore we have \beq\label{ido}
\eta= {\rm Re}( \lambda_{\pm}\mp\12 q), \ \hat{\eta}= \lambda_{\pm}\mp \12 q.
\eeq  
In this situation we will call $\eta$ the {\em real covariance} of the state $\omega$, to distinguish it from the complex covariances $\lambda_{\pm}$.

 \begin{remark} the state $\omega$ is of course  uniquely determined by either $\lambda_{+}$ or $\lambda_{-}$, but later  conditions on a state $\omega$ look nicer  when formulated in terms of the pair of  covariances $\lambda_{\pm}$.
Note that  $\lambda_{-}$ is usually called the {\em  charge density} associated to  $\omega$.
\end{remark}

The following  propositions are the analogues of Props. \ref{prop:defomega}, \ref{ista}, \ref{istaa}. We sketch their proofs for the reader's convenience.
\begin{proposition}\label{isto}
 Let $\lambda_{\pm}\in L_{\rm s}(\cY, \cY^{*})$. Then the following are equivalent:
 \ben
 \item $\lambda_{\pm}$ are  the covariances of a gauge-invariant quasi-free state on $\mathcal{A}(\cY, q)$,
 \item $\lambda_{\pm}\geq 0$ and $\lambda_{+}- \lambda_{-}=q$.
 \een 
\end{proposition}
\proof 
Since $\omega$ is gauge-invariant we have 
\[
\ii\in O(\cY_{\rr}, \eta)\cap Sp(\cY_{\rr}, {\rm Re}\sigma)= O(\cY_{\rr}, \eta)\cap O(\cY_{\rr}, {\rm Re}q).
\]
From this fact  and (\ref{ido}) we deduce that $\eta\geq 0\Leftrightarrow\lambda_{+}\geq \12 q$, and that the  second condition in Prop. \ref{prop:defomega} (with $\sigma$ replaced by $\Re\sigma$) is equivalent to 
\[
\pm q\leq 2\lambda_{+} -q\ \Leftrightarrow \ \lambda_{\pm}\geq 0.
\]
This completes the proof of the proposition. \qed
\begin{proposition}\label{istu}
Let $\lambda_{\pm}\in L_{\rm s}(\cY, \cY^{*})$. Then the following are equivalent:
 \ben
 \item $\lambda_{\pm}$ are  the covariances of a {\em pure} gauge-invariant quasi-free state on $\mathcal{A}(\cY, q)$,
 \item there exists  an involution $\kappa\in U(\cY, q)$  such that $q\kappa\geq 0$ and $\lambda_{\pm}= \12 q(\kappa\pm \one)$.
 \item $\lambda_{\pm}\geq \pm\12 q$, $ \lambda_{\pm} q^{-1}\lambda_{\pm}= \pm \lambda_{\pm}$, $\lambda_{+}- \lambda_{-}=q$.
 \een
\end{proposition}
\proof   By Prop. \ref{ista} the state $\omega$ is pure iff there exists  an anti-involution $\jj\in Sp(\cY_{\rr}, {\rm Re}\sigma)$ such that \beq\label{toto}
2\eta= ({\rm Re}\sigma) \jj.
\eeq  Since  $\ii\in O(\cY_{\rr}, \eta)\cap Sp(\cY_{\rr}, {\rm Re}\sigma)$ we obtain  that $\jj\in U(\cY, q)$, i.e. $\jj$ is $\cc-$linear and pseudo-unitary for $q$. From (\ref{toto}) we then get that $2\lambda_{+}-q= \sigma\jj$. Setting $\kappa= -\ii\jj$ we see that $\kappa\in U(\cY, q)$ and $\lambda_{+}= \12 q(\one + \kappa)$.  
Therefore (1) is equivalent to
\[
(4)\ \lambda_{+}\geq 0, \ \lambda_{+}\geq q, \ \lambda_{+}= \12 q(\one + \kappa), \ \kappa^{2}= \one,  \kappa\in U(\cY, q).
\]
(4) clearly implies (2). Let us prove the converse implication.  Set $P_{\pm}:= \12(\one \pm \kappa)$. Clearly $P_{\pm}$ are projections with $P_{\pm}^{*}q= qP_{\pm}$,  $\kappa P_{\pm}= \pm P_{\pm}$, and 
\[
\lambda_{+}\geq 0, \ \lambda_{+}\geq q \Leftrightarrow \pm qP_{\pm}\geq 0.
 \]
 Now we have
 \[
 qP_{\pm}= qP_{\pm}^{2}= P_{\pm}^{*}qP_{\pm}= \pm P_{\pm}^{*}q\kappa P_{\pm},
\]
which completes the proof since $q\kappa\geq 0$.  The fact that $(2)$ and $(3)$ are equivalent  is an easy computation. \qed

\begin{proposition}
 Let $\lambda_{\pm}$, $\tilde{\lambda}_{\pm}$ be the covariances of two pure, gauge-invariant quasi-free states on $\mathcal{A}(\cY, q)$. Then there exists $r\in U(\cY, q)$ such that $\tilde{\lambda}_{\pm}= r^{*}\lambda_{\pm}r$.
\end{proposition}
\proof We introduce the real covariances $\eta$, $\tilde\eta$. By Prop. \ref{istaa} there exists $r\in Sp(\cY_{\rr}, {\rm Re}\sigma)$ with $\tilde{\eta}= r^{\t}\eta r$. Using the gauge-invariance of the two states  we obtain that $r\ii= \ii r$, hence $r\in U(\cY, q)$.  From this we easily obtain the proposition.  \qed
\subsection{Charge reversal}
\begin{definition}
 Let $(\cY, \sigma)$ a complex symplectic space. \ben 
 \item a map $\chi\in L(\cY_{\rr})$ is called a {\em charge reversal} if $\chi^{2}= \one$ or $\chi^{2}= -\one$ and 
 \[
(\chi y_{1}| \sigma \chi y_{2})= (y_{2}| \sigma y_{1}), \ y_{1}, y_{2}\in\cY.
\]
\item 
a gauge invariant quasi-free state $\omega$ on $\mathcal{A}(\cY, q)$ is {\em invariant }under  the charge reversal $\chi$ if
\[
\omega(\phi(\chi y_{1})\phi(\chi y_{2}))= \omega(\phi(y_{1})\phi(y_{2})), \ y_{1}, y_{2}\in \cY,
\]
or equivalently:
\[
\omega(\psi(\chi y_{1})\psi^{*}(\chi y_{2}))= \omega(\psi^{*}(y_{1})\psi(y_{2})),  \ y_{1}, y_{2}\in \cY.
\]
\een 
\end{definition}
Note that  a charge reversal is automatically anti-linear. The last condition above can be rephrased as
\beq\label{reversal}
 \chi^{*}\lambda_{\pm} \chi= \lambda_{\mp}.
\eeq
\subsection{Complexification of   bosonic quasi-free states}\label{complix}
\subsubsection{Neutral case}
Let now $(\cX, \sigma)$ be a real symplectic space.  We   equip $\cc\cX$ with $\sigma_{\cc}$, obtaining a complex symplectic space.  We set as in Subsect. \ref{sec0.2} $q= \i \sigma_{\cc}$. The canonical complex conjugation on $\cc\cX$ is a charge reversal on $(\cc\cX, \sigma_{\cc})$.

Clearly $((\cc\cX)_{\rr}, \Re \sigma_{\cc})$ is isomorphic to $(\cX\oplus \cX, \sigma\oplus \sigma)$ as  real symplectic spaces.
If $\omega$ is a quasi-free state on $(\cX, \sigma)$ with covariance $\eta$, then we can consider the quasi-free state $\tilde{\omega}$ on $\mathcal{A}((\cc\cX)_{\rr}, \Re \sigma_{\cc})$ with covariance $\Re \eta_{\cc}$.

It is easy to see that $\tilde{\omega}$ is gauge-invariant with  covariances $\lambda_{\pm}$  equal to
\[
\lambda_{\pm}= \eta_{\cc}\pm \12 q.
\]
Moreover $\tilde{\omega}$ is invariant under charge reversal.

Therefore by complexifying a quasi-free state $\omega$ on a real symplectic space $(\cX, \sigma)$, we obtain a gauge-invariant quasi-free state $\tilde{\omega}$ on $\mathcal{A}(\cc\cX, \sigma_{\cc})$. It follows that, possibly after complexifying the real symplectic space $(\cX, \sigma)$, one can always restrict the discussion to gauge-invariant  quasi-free states.

\subsubsection{Charged case}\label{chargi}
Assume now that $(\cX, \sigma)$ is a complex symplectic space, with complex structure $\ii$ and  $\omega$ be a gauge-invariant quasi-free state on $\mathcal{A}(\cX, \sigma)$, with real covariance $\eta$.

We can apply the above procedure to  the real symplectic space $(\cX_{\rr}, \Re \sigma)$.  Then $\cc \cX$ has two complex structures, the canonical one $\i$ and $\ii_{\cc}$,  (the complexification of $\ii$). As is well known, (see e.g. \cite[Sect. 1.3.6]{DG}) $\cc\cX$ splits as the direct sum 
\[
\cc\cX= \cZ\oplus \overline{\cZ},
\]
where
\[
 \cZ:= \Ker(\ii_{\cc} -\i ), \ \overline{\cZ}:= \Ker(\ii_{\cc}+\i)
\]
are called the {\em holomorphic} resp. {\em anti-holomorphic subspaces} of $\cc\cX$.   Note that the natural conjugation on $\cc\cX$ maps bijectively $\cZ$ onto $\overline{\cZ}$. Then (see \cite[Sect. 17.1.2]{DG}) the sequilinear extensions $\eta_{\cc}$ and $(\Re \sigma)_{\cc}$ are reduced w.r.t. the direct sum $\cZ\oplus \overline{\cZ}$, i.e.:
\[
(\Re \sigma)_{\cc}:= \mat{\sigma_{\cZ}}{0}{0}{\overline{\sigma}_{\cZ}}, \ \eta_{\cc}:= \mat{\eta_{\cZ}}{0}{0}{\overline{\eta}_{\cZ}},
\]
where $\sigma_{\cZ}\in L_{\rm a}(\cZ, \cZ^{*})$, $\eta_{\cZ}\in L_{\rm s}(\cZ, \cZ^{*})$ and if $a\in L(\cZ, \cZ^{*})$ we define $\overline{a}\in L(\overline{\cZ}, \overline{\cZ}^{*})$ by $(\overline{z}_{1}| \overline{a}\overline{z}_{1}):= \overline{(z_{1}| a z_{2})}$.

Let us  remark that one can identify $\cX$ with $\cZ$ by the ($\cc-$linear) map
\[
T: \cX\ni x\mapsto Tx= \frac{1}{\sqrt{2}}(x- \i \ii x)\in \cZ, 
\]
and also $\overline{\cX}$ with $\overline{\cZ}$ by $\overline{T}$.
Under these identifications,  $\sigma_{\cZ}$  is identified with $\sigma$,  $\eta_{\cZ}$ with $\hat{\eta}$, where $\hat{\eta}$ is the $\ii-$sesquilinear extension of $\eta$ defined in (\ref{tati}). It follows then from (\ref{ido}) that   the sesquilinear form $\eta_{\cc}+ \frac{\i}{2}(\Re \sigma)_{\cc}$ acting on $\cc\cX$ is identified with
\[
\mat{\lambda_{+}}{0}{0}{\overline{\lambda_{-}}}, \hbox{ acting on }\cX\oplus \overline{\cX}.
\]

\section{Hadamard  states}\label{sec22}
\subsection{Klein-Gordon equations on a globally hyperbolic space time}\label{sec22.0}
\subsubsection{Notation}
 Let $M$ be a smooth manifold.  As usual  $\cE(M)$ is the space of smooth, (complex valued) functions on $M$,  $\cD(M)\subset \cE(M)$ the space of smooth  compactly supported functions on $M$, $\cD'(M)$ the space of distributions on $M$ and $\cE'(M)$ the space of compactly supported distributions.   We denote by $\langle u|v\rangle$, for $u\in \cD(M)$ (resp. $\cE(M)$), $v\in \cD'(M)$ (resp. $\cE'(M)$) the (bilinear) duality bracket.
 \subsubsection{Klein-Gordon equations}
 Assume $M$ is equipped with a Lorentzian metric $g=g_{\mu\nu}dx^{\mu}dx^{\nu}$ such that $(M, g)$  is a globally hyperbolic space-time.  We use the convention $(-, +,\cdots ,+)$ for the signature.

We use the notations \[
|g|:= {\rm det}[g_{\mu\nu}], \quad [g^{\mu\nu}]:= [g_{\mu\nu}]^{-1}, \quad dv:= |g|^{\12}dx.
\]

If $S$ is a Cauchy hypersurface, we denote by $n^{\nu}$ the unit  future directed normal vector field to $S$ (after choosing  a time orientation), and by $ds$ the surface measure on $S$ obtained from $dv$.

We fix a smooth vector potential $A_{\mu}(x) dx^{\mu}$ and a smooth function $\rho: M\to \rr$. 
The associated Klein-Gordon operator is:
\beq\label{e0.00}
P(x, D_{x})= |g|^{-\12}(\p_{\mu}+ \i A_{\mu})|g|^{\12}g^{\mu\nu} (\p_{\nu}+ \i A_{\nu})+\rho.
\eeq
We equip $\cD(M)$ with the scalar product
\[
(u_{1}|u_{2})= \int_{M} \overline{u}_{1}u_{2} dv,
\]
so that $P(x, D_{x})$ is formally selfadjoint. We denote by 
$E_{\pm}$ the {\em retarded/advanced fundamental solutions} of $P(x, D_{x})$,
 and by $E= E_{+}- E_{-}$ the {\em Pauli-Jordan commutator function} Recall that $ E_{\pm}^{*}= E_{\mp}$ hence $E= - E^{*}$.

A function $u$ on $M$ is called {\em space-compact} if the intersection of $\supp\, u$ with any Cauchy hypersurface of $M$ is compact. The space of smooth space-compact functions will be denoted by $\cinf_{\rm sc}(M)$.

We denote by ${\rm Sol}_{\rm sc}(P)\subset \cinf(M)$ the space of   smooth space-compact solutions of 
\[
\kg \quad P(x, D_{x})\phi=0.
\]
One has (see e.g. \cite{BGP}):
\beq\label{e0.1}
 E \cD(M)={\rm Sol}_{\rm sc}(P), \ \Ker E= P\cD(M).
\eeq
Moreover if  we  fix a Cauchy hypersurface $S$ and set
\[
\begin{array}{rcl}
\rho:& {\rm Sol}_{\rm sc}(P)&\to \cD(S)\oplus \cD(S)\\[2mm]
&\phi&\mapsto (\phi_{\mid S}, \i^{-1}n^{\mu}(\nabla_{\mu}+ \i A_{\mu})\phi_{\mid S})=: (\rho_{0}\phi, \rho_{1}\phi),
\end{array}
\]
then $\rho: {\rm Sol}_{\rm sc}(P)\to \cD(S)\oplus \cD(S)$ is bijective (see e.g. \cite{BGP}).

Let $\varsigma$ be the sesquilinear form on $\cD(M)/P\cD(M)$ defined by 
\[
([u]|\varsigma[v]):=\bra \overline{u}, E v\ket =\bra E , \overline{u}\otimes v\ket, \quad u,v\in\cD(M).
\]
By construction $(\cD(M)/P\cD(M),\varsigma)$ is a complex symplectic space.  Setting also 
\[
(f|\sigma g):= -\i\int_{S}( \overline{f_{0}}g_{1}+ \overline{f_{1}}g_{0})ds, \  f,g\in\cD(S)\oplus \cD(S),
\]
we have
\beq\label{e0.2}
([u]|\varsigma [v])= (\rho \circ E u|\sigma\rho \circ E v),
\eeq
i.e. 
\[
\rho\circ E: (\cD(M)/P\cD(M), \varsigma)\to (\cD(S)\oplus \cD(S), \sigma)
\]
is a symplectomorphism. 
%
%
%

\subsection{The wave front set}\label{sec-1}
Let $M$ be a smooth manifold,  $\co{M}$ its cotangent bundle of $M$. The zero section  of $\co{M}$  will be denoted by $Z$.   

\subsubsection{Operations on conic sets}\label{sec-1.1}
A set $\Gamma\subset \coo{M}$ is {\em conic} if 
\[
(x, \xi)\in \Gamma\Rightarrow (x, t \xi)\in \Gamma, \ \forall \ t>0.
\]
If  $\Gamma\subset \coo{M}$ is conic, we set:
\[
\overline{\Gamma}:=\{(x, -\xi): \  (x, \xi)\in \Gamma\}. 
\]
Let $M_{i}$, $i=1,2$ be two manifolds, $Z_{i}$ the zero section of $\co{M_{i}}$ and $\Gamma\subset \coo{(M_{1}\times M_{2})}$ be a conic set.   The elements of $\coo{(M_{1}\times M_{2})}$
 will be denoted by $(x_{1}, \xi_{1}, x_{2}, \xi_{2})$, which allows to 
 consider $\Gamma$ as a relation  between
 $\T^{*}M_{2}$ and $T^{*}M_{1}$, still denoted by $\Gamma$. Clearly $\Gamma$ maps conic sets into conic sets.
 We set:
 \[
\begin{array}{l}
\Gamma':= \{(x_{1}, \xi_{1}, x_{2}, - \xi_{2}) :\ (x_{1}, \xi_{1}, x_{2}, \xi_{2})\in \Gamma\}\subset \coo{(M_{1}\times M_{2})},\\[2mm]
{\rm Exch}(\Gamma):= \{(x_{2}, \xi_{2}, x_{1}, \xi_{1}) : \ (x_{1}, \xi_{1}, x_{2}, \xi_{2})\in \Gamma\}\subset \coo{(M_{2}\times M_{1})},\\[2mm]
_{M_{1}}\!\Gamma:=\{(x_{1}, \xi_{1}) : \ \exists \ x_{2}\hbox{ such that } (x_{1}, \xi_{1}, x_{2},0)\in \Gamma\}= \Gamma(Z_{2})\subset \coo{M_{1}}_{1},\\[2mm]
\Gamma\!_{M_{2}}:=\{(x_{2}, \xi_{2}) : \ \exists \ x_{1}\hbox{ such that } (x_{1}, 0, x_{2},\xi_{2})\in \Gamma\}= \Gamma^{-1}(Z_{1})\subset\coo{M_{2}}_{2}.
\end{array}
\]

\subsubsection{Properties of the wave front set}\label{sec-1.2}
Recall that if $u\in \cD'(M)$ then the wave front set $\WF(u)$ is a conic subset of $\coo{M}$. We refer to \cite{hoermander1} for the exact definition and the proof of the following basic properties:

\medskip

\ben
\item {\bf Complex conjugation}: if $u\in \cD'(M)$ then $\WF(\overline{u})= \overline{\WF(u)}$.

\item {\bf Restriction to a sub-manifold}: let $S\subset M$ a sub-manifold. The {\em co-normal bundle} to $S$ in $M$ is:
\[
T_{S}^{*}M:=\{(x, \xi)\in \coo{M} : \ x\in S, \ \xi\cdot v=0\ \forall v\in T_{x}S\}.
\]
If $u\in\cD'(M)$, the restriction $u_{|S}$ of $u$ to $S$ is well defined if
$\WF(u)\cap T_{S}^{*}M=\emptyset$.  One has
\[
\WF(u_{|S})\subset \{(x, \xi_{|T_{x}S}) : \ x\in S, \ (x, \xi)\in \WF(u)\}.
\]
\item {\bf Kernels}: let $K: \cD(M_{2})\to \cD'(M_{1})$ be linear continuous and denote  by $K(x_{1}, x_{2})\in \cD'(M_{1}\times M_{2})$ its distributional kernel. Then $Ku$ is well defined for $u\in \cE'(M_{2})$ if $\WF(u)\cap \WF(K)'\!_{M_{2}}=\emptyset$ and in such case
\[
\WF(Ku)\subset \, _{M_{1}}\!\WF(K)\cup  \WF(K)'\circ \WF(u).
\]
\item {\bf Composition}: let  $K_{1}\in \cD'(M_{1}\times M_{2})$, $K_{2}\in \cD'(M_{2}\times M_{3})$, where $K_{2}$ is properly supported, i.e. the projection:
$\supp\,K_{2}\to M_{2}$ is proper. Then $K_{1}\circ K_{2}$ is well defined if
\[
\WF(K_{1})'\!_{M_{2}}\, \cap \, _{M_{2}}\!\WF(K_{2})'= \emptyset,
\] 
and in such case
\[
\WF(K_{1}\circ K_{2})'\subset \WF(K_{1})'\circ \WF(K_{2})'\ \cup \ _{M_{1}}\!\WF(K_{1})'\times Z_{3}\ \cup \ Z_{1}\times \WF(K_{2})'\!_{M_{3}}.
\]
\item {\bf Adjoint:} let us denote by $K^{*}$ the adjoint of $K$ with respect to any smooth non-vanishing density $dx$ on $M$. Then
\[
\WF(K^{*})'= {\rm Exch}(\WF (K)').
\]
\een

\subsection{Distinguished parametrices and microlocal spectrum condition}
\subsubsection{Distinguished parametrices}
Let us recall basic elements of the theory of distinguished parametrices of Duistermaat and H\"ormander  \cite{duistermaat-hormander}  for the case of the Klein-Gordon operator $P(x,D)$. The {\em characteristic manifold} of $P(x,D)$ is
\[
\cN:=\{ (x,\xi)\in \coo{M} : \ p(x,\xi)=0 \},
\]
where $p(x,\xi)=g^{\mu\nu}(x)\xi_{\mu}\xi_{\nu}$ is the principal symbol of $P(x,D)$.

We use the notation $X=(x,\xi)$ for points in $\coo{M}$. We write $X_1\sim X_2$ if $X_1=(x_1,\xi_1)$ and $X_2=(x_2,\xi_2)$ are in $\cN$ and $X_1$ and $X_2$ are on the same Hamiltonian curve for $p$.

Let us denote  by $V_{x\pm}\subset T_x M$ for $x\in M$, the open future/past light cones and $V_{x\pm}^*$ the dual cones
\[
V^{*}_{x\pm}:=\{ \xi\in T^*_{x}M : \ \xi\cdot v >0, \ \forall v\in V_{x\pm}, \ v\neq 0 \}.
\]
The set $\cN$ has two connected components invariant under the Hamiltonian flow of $p$, namely:
\[
\cN_{\pm}:=\{ X\in \cN : \ \xi\in V^*_{x\pm}\}.
\]

Recall that $E_{\pm}$ denote respectively the retarded and advanced fundamental solution. We denote respectively $E_{\rm  F}$, $E_{\bar{\rm  F}}$ the Feynman and anti-Feynman parametrix. The theory of Duistermaat and H\"ormander provides among others a description of the wave front sets of the parametrices $E_{\pm}$, $E_{\rm  F}$, $E_{\bar{\rm  F}}$ and establishes their uniqueness up to smooth functions. 
The proof of the next lemma can be found for instance in \cite[Thm. 2.29]{junker}. 

\begin{lemma}\label{lem:raff}We have:
\begin{eqnarray*}
(1) & \WF(E)'=\{(X_{1}, X_{2})\in \cN\times \cN : \  \ X_{1}\sim X_{2}\},\\
(2) & \WF(E_{\rm +}-E_{\rm F})'=\{(X_{1}, X_{2})\in \cN_{-}\times \cN_{-} : \  \ X_{1}\sim X_{2}\},\\
(3) & \WF(E_{\rm -}-E_{\rm F})'=\{(X_{1}, X_{2})\in \cN_{+}\times \cN_{+} : \  \ X_{1}\sim X_{2}\}.
\end{eqnarray*}
\end{lemma}
\subsubsection{Microlocal spectrum condition}
We are now ready to formulate the microlocal spectrum condition.

Let us  fix a gauge-invariant, quasi-free state $\omega$ on $\mathcal{A}(\cD(S)\oplus \cD(S), \sigma)$. We denote by $\eta$ its real covariance and $\lambda_{\pm}$ its two complex covariances (see Subsect. \ref{sec0.2}).   We assume that 
\[
\eta: \cD(S)\oplus \cD(S)\to \cD'(S)\oplus \cD'(S)
\]
(and hence $\lambda_{\pm}$) is continuous. Note that $\lambda_{\pm}$ are $\cc-$linear, while $\eta$ is only $\rr-$linear.

We associate to $\eta$, $\lambda_{\pm}$ linear operators  $H$, $\Lambda_{\pm}\in L(\cD(M), \cD'(M))$ as follows:
\beq\label{defino}
\begin{array}{rl}
\langle u_{1}| H u_{2}\rangle& := (\rho\circ E )u_{1}\dito \eta  (\rho\circ E) u_{2}, \\[2mm]
 (u_{1}| \Lambda_{\pm} u_{2})&:= \left( (\rho\circ E )u_{1}| \lambda_{\pm}  (\rho\circ E) u_{2}\right).
\end{array}
\eeq
Note that $\Lambda_{\pm}$ are $\cc-$linear, while $H$ is only $\rr-$linear.  From Subsects. \ref{sec0.2}, \ref{sec22.0} we have
$\Lambda_{+}- \Lambda_{-}=\i E$.
\begin{definition}\label{def2}
 Let $\Lambda_{\pm}: \cD(M)\to \cD'(M)$ be as in (\ref{defino}). Then $\Lambda_{\pm}$ satisfies the {\em microlocal spectrum condition} if
\beq\label{eq:msc}\tag{$\mu$sc}
\WF(\Lambda_{\pm})'\subset \{(X_{1}, X_{2})\in \cN_{\pm}\times \cN_{\pm} : \  \ X_{1}\sim X_{2}\}.
\eeq
A gauge invariant quasi-free state is a {\em Hadamard state} if its complex covariances $\Lambda_{\pm}$ satisfy the microlocal spectrum condition.
\end{definition}

\begin{remark}
  The above form of the Hadamard condition for charged fields seems to have been first noticed by Hollands \cite{hollands} for Dirac fields. Let us explain its relationship with the standard form in \cite{radzikowski}: in the standard form, symplectic spaces (real or complex) are considered as real symplectic spaces, and a state is  labelled by its real covariance. The standard Hadamard condition takes then the form
  \beq\label{radiko}
  \WF(H_{\cc}+ \frac{\i}{2}(\Re E)_{\cc})'\subset  \{(X_{1}, X_{2})\in \cN_{+}\times \cN_{+} : \  \ X_{1}\sim X_{2}\}.
\eeq
Note that if we consider a charged Klein-Gordon equation, $H_{\cc}+ \frac{\i}{2}(\Re E)_{\cc}$ is actually a $2\times 2$ matrix of distributions on $M\times M$, its wave front set being then the union of the wave front set of its entries.

We can apply  the discussion of Subsubsect. \ref{chargi} and identify $H_{\cc}+ \frac{\i}{2}(\Re E)_{\cc}$ with the matrix $\mat{\Lambda_{+}}{0}{0}{\overline{\Lambda_{-}}}$. Using that $\WF(\overline{\Lambda_{-}})'= \overline{\WF(\Lambda_{-})'}$, we see that (\ref{radiko}) is equivalent to (\ref{eq:msc}).

 One can also show that the inclusion in (\ref{eq:msc}) and (\ref{radiko}) can be replaced by an equality.
\end{remark}
\begin{remark}
 Assume now that we consider a neutral Klein-Gordon equation, i.e. that $A_{\mu}=0$. Then the Klein-Gordon operator $P$ is real, and the map $u\mapsto \overline{u}$  is a charge reversal of the symplectic spaces $({\rm Sol}_{\rm sc}(P), \sigma)$, $(\cD(M)/P\cD(M), \varsigma)$.  It follows then from (\ref{reversal}) that $\overline{\Lambda}_{\pm}= \Lambda_{\mp}$, so each of the two conditions in (\ref{eq:msc}) implies the other.
\end{remark}

\section{Background on pseudo-differential  calculus}\label{sec1}
\subsection{Notation}\label{sec1.0}
- If $f:\rr_{t}\times \rr^{n}_{x}\to \cc$ is a function, and $t\in \rr$ we denote by $f(t)$ the function:
\[
f(t): \ \rr^{n}\ni x\mapsto f(t, x)\in \cc.
\]

- We denote by $C^{\infty}_{\rm bd}(\rr^{n})$ the space of smooth functions on $\rr^{n}$ uniformly bounded with all derivatives. We equip $C^{\infty}_{\rm bd}(\rr^{n})$ with its canonical Fr\'echet space structure. We denote by $H^{m}(\rr^{d})$  the Sobolev space of order $m\in \rr$.

- We denote by $\cS(\rr^{d})$, resp. $\cS'(\rr^{d})$ the space of Schwartz functions, resp. distributions on $\rr^{d}$.

- We set
\[
 \cH(\rr^{d}):=\cap_{m\in \rr}H^{m}(\rr^{d}), \ \cH'(\rr^{d}):=\cup_{m\in \rr}H^{m}(\rr^{d}),
 \]
  equipped with their canonical topologies.

- If $E, F$ are two topological vector spaces, we write $A: E\to F$ if $A$ is linear continuous from $E$ to $F$.

- We set $D_{x}= \i^{-1}\p_{x}$, $\langle x\rangle= (1+x^{2})^{\12}$, $x\in \rr^{d}$.
\subsection{Symbol classes}\label{sec1.1}
We denote by $S^{m}\!(\rr^{2d})$, $m\in \rr$ the  symbol class 
\beq\label{e1.1}
a\in S^{m}\!(\rr^{2d}) \ \ \hbox{if} \ \ \p_{x}^{\alpha}\p_{k}^{\beta}a(x, k)\in O(\langle k\rangle^{m-|\beta|}), \ \alpha, \beta\in \nn^{d}.
\eeq
Similarly we will denote by $S^{m}\!(\rr)$ the class
\begin{equation}
\label{e1.1b}
f\in S^{m}\!(\rr) \hbox{ if }\p_{\lambda}^{\alpha}f(\lambda)\in O(\langle \lambda\rangle^{m-\alpha}), \ \alpha\in \nn.
\end{equation}
We denote by $S^{m}_{\rm h}\!(\rr^{2d})$ the subspace of $S^{m}\!(\rr^{2d})$ of symbols homogeneous of degree $m$ in the $k$ variable, (outside a neighborhood of the origin) :
\beq\label{e1.2}
a\in S^{m}_{\rm h}\!(\rr^{2d})\hbox{ if }a\in S^{m}\!(\rr^{2d})\hbox{ and }a(x, \lambda k)= \lambda^{m}a(x, k), \ \lambda\geq 1, \  |k|\geq 1.
\eeq
We set
\[
S^{-\infty}(\rr^{2d}):= \bigcap_{m\in \rr}S^{m}(\rr^{2d}).
\]
If $a_{m-k}\in S^{m-k}\!(\rr^{2d})$ for $k\in \nn$ and $a\in S^{m}\!(\rr^{2d})$ we write 
\[
a\sim \sum_{k=0}^{\infty}a_{m-k}
\]
if
\begin{equation}
\label{e1.3}
a-\sum_{k=0}^{n}a_{m-k}\in S^{m-n-1}\!(\rr^{2d}), \ \forall n\in \nn.
\end{equation}
Note that if $a_{m-k}\in S^{m-k}\!(\rr^{2d})$ for $k\in \nn$, then it is well-known that there exists $a\in S^{m}(\rr^{2d})$, unique modulo $S^{-\infty}(\rr^{2d})$, such that $a\sim \sum_{k=0}^{\infty}a_{m-k}$.

We say that a symbol $a\in S^{m}(\rr^{2d})$ is {\em poly-homogeneous} if $a\sim\sum_{k=0}^{\infty}a_{m-k}$ for $a_{m-k}\in S^{m-k}_{\rm h}\!(\rr^{2d})$.The symbols  $a_{m-k}$ are then   clearly unique modulo $S^{-\infty}(\rr^{2d})$.
  
The   subspace of {\em poly-homogeneous symbols} of degree $m$ will be denoted by $S^{m}_{\rm ph}\!(\rr^{2d})$.
The space $S_{\rm ph}^{m}(\rr)\subset S^{m}(\rr)$ is defined similarly.

We will often write  $S^{m}_{({\rm ph})}$  for $S^{m}_{({\rm ph})}\!(\rr^{2d})$. We equip  $S^{m}_{({\rm ph})}\!(\rr^{2d})$ with the Fr\'echet space topology given by the semi-norms:
\[
\| a\|_{m, N}:= \sup_{|\alpha|+ |\beta|\leq N}| \langle k\rangle^{-m+ |\beta|}\p_{x}^{\alpha}
\p_{k}^{\beta}a|.
\]
We set
\[
S^{\infty}_{({\rm ph})}(\rr^{2d}):=\bigcup_{m\in \rr}S^{m}_{({\rm ph})}(\rr^{2d}).
\]
\subsection{Principal symbol and characteristic set}\label{sec1.2}

The {\em principal symbol} of $a\in S^{m}$, denoted by $\sigma_{\rm pr}(a)$ is the equivalence class $a+ S^{m-1}$ in $S^{m}/S^{m-1}$. If $a\in S^{m}_{\rm ph}$ then $a+ S^{m-1}$ has a unique representative in $S^{m}_{\rm h}$, namely the function $a_{m}$ in (\ref{e1.3}). 
Therefore  in this case the principal symbol is a  function on $ \rr^{2d}$, homogeneous of degree $m$ in $k$.

The {\em characteristic set} of $a\in S^{m}_{\rm ph}$ is defined as
\begin{equation}
\label{e1.4}
{\rm Char}(a):= \{(x, k)\in\cooo{\rr^{d}} : \  a_{m}(x, k)= 0\},
\end{equation}
it is clearly conic  in the $k$ variable.

A symbol $a\in S^{m}\!(\rr^{2d})$ is {\em elliptic} if  there exists $C, R>0$ such that
\[
 |a(x, k)|\geq C \langle k\rangle^{m}, \ |k|\geq R.
\]
Clearly $a\in S^{m}_{\rm ph}\!(\rr^{2d})$ is elliptic iff ${\rm Char}(a)= \emptyset$.

\subsection{Pseudo-differential operators}\label{sec1.3}

In this subsection we collect some well-known results about pseudo-differential  calculus. 

For $a\in S^{m}\!(\rr^{2d})$, we denote by  $\Opw(a)$ the {\em Weyl quantization} of $a$ defined by:
\[
\Opw(a)u(x)=  a^{\rm w}(x, D)u(x):=(2\pi)^{-d}\iint\e^{\i
(x-y)k}a(\frac{x+ y}{2}, k)u(y)\d y\d k.
\]
One has $\Opw(a): \cH(\rr^{d})\to \cH(\rr^{d})$  and
\[
\Opw(a)^{*}= \Opw(\overline{a}),
\]
 hence $\Opw (a):  \cH'(\rr^{d})\to \cH'(\rr^{d})$. 

We denote by  $\Psi^{m}_{({\rm ph})}(\rr^{d})$ the space $\Opw (S^{m}_{({\rm ph})}\!(\rr^{2d}))$ and set
\[
\Psi_{({\rm ph})}^{\infty}(\rr^{d})= \bigcup_{m\in \rr}\Psi^{m}_{({\rm ph})}(\rr^{d}), \ \Psi^{-\infty}(\rr^{d})= \bigcap_{m\in \rr}\Psi^{m}(\rr^{d}).
\]
We will often write $\Psi^{m}_{({\rm ph})}$ instead of $\Psi^{m}_{({\rm ph})}(\rr^{d})$. We will equip $\Psi^{m}_{({\rm ph})}(\rr^{d})$ with the  Fr\'{e}chet space topology obtained from the topology of $S^{m}(\rr^{2d})$.

If $a= a^{\rm w}\!(x, D_{x})\in \Psi^{m}_{\rm ph}(\rr^{d})$ the $m-$homogeneous   function 
$\sigma_{\rm pr}(a)=:a_{m}(x, k)$ is called the {\em principal symbol} of $a$.

Let $s,m\in \rr$. Then   the map
\beq\label{e0.0}
S^{m}\!(\rr^{d})\ni a\mapsto \Opw(a)\in B(H^{s}(\rr^{d}), H^{s-m}(\rr^{d}))
\eeq
is continuous.

 An operator $\Opw\!(a)\in \Psi^{m}(\rr^{2d})$ is {\em elliptic} if its symbol $a(x, k)$ is elliptic in $S^{m}\!(\rr^{2d})$.  If $a\in \Psi^{m}$ is elliptic 
 then there exists $b\in \Psi^{-m}$, unique modulo $\Psi^{-\infty}$ such that $ab=ba=\one$ modulo $\Psi^{-\infty}$. Such an operator $b$ is called a {\em pseudo-inverse} or a {\em parametrix} of $a$. We will denote it by $b=:a^{(-1)}$. As a typical example  $\one + b$ for $b\in \Psi^{-m}$, $m>0$ is elliptic in $\Psi^{0}$.

The following lemma is proven in  \cite[Prop. A.1.1, A.1.2]{S} .

\begin{lemma}\label{1.0}
 Let $u\in \cD'(\rr^{d})$, $(x_{0}, k_{0})\in \cooo{\rr^{d}}$. Then $(x_{0}, k_{0})\not \in \WF(u)$ iff there exists $\chi\in \coinf(\rr^{d})$  and $a\in S^{0}_{\rm ph}\!(\rr^{2d})$ with $\chi(x_{0})\neq 0$, $(x_{0}, k_{0})\not\in {\rm Char}(a)$ and $\Opw\!(a)\chi u\in \cS(\rr^{d})$.
\end{lemma}
\subsection{Functional calculus for pseudo-differential  operators}\label{sec1.4}

We now recall various well-known results about functional calculus for pseudo-differential  operators.
\begin{proposition}\label{1.1}
Let $a\in \Psi^{m}(\rr^{d})$  for $m\geq 0$ be elliptic in $\Psi^{m}(\rr^{d})$ and symmetric on $\cS(\rr^{d})$. Then:
\ben \item $a$ is selfadjoint on $H^{m}(\rr^{d})$,
\item  Denote by $\rho(a)$ the resolvent set of $a$, with domain $H^{m}(\rr^{d})$. Then for $z\in \rho(a)$ $(z-a)^{-1}\in \Psi^{-m}(\rr^{d})$,
\item if $f\in S^{p}(\rr)$, $p\in \rr$,  then $f(a)$, defined by the functional calculus, belongs to $\Psi^{mp}(\rr^{d})$. 
\item if  $f$ is elliptic in $S^{p}(\rr)$ then $\sigma_{\rm pr}(f(a))= f(\sigma_{\rm pr}(a))$ mod $S^{mp-1}(\rr^{2d})$.
\item if $a\in \Psi^{m}_{\rm ph}(\rr^{d})$ and $f\in S^{p}_{\rm ph}(\rr)$, then $f(a)\in \Psi^{mp}_{\rm ph}(\rr^{d})$, and if $f\in S^{p}_{\rm ph}(\rr)$ is elliptic, then $\sigma_{\rm pr}(f(a))= f_{p}(\sigma_{\rm pr}(a))$, where $f_{p}\in S^{p}_{\rm h}(\rr)$ is the principal symbol of $f$.
\een
\end{proposition}
\proof 
We refer the reader for example to \cite[Thm. 5.4]{Ro}, \cite[Corr. 4.5]{Bo} for the proof of similar statements.   Statement (1) follows from the fact that $a+ \i \lambda \one$ maps $H^{m}(\rr^{d})$ bijectively onto $L^{2}(\rr^{d})$ for $|\lambda|$ large enough. 

To prove statement (2), the most direct way is to use the Beals criterion (see e.g. \cite{Bo}), which characterizes pseudo-differential  operators by properties of the multi-commutators with the operators $x_{i}$, $D_{j}$:  an operator $a$ belongs to $\Psi^{m}(\rr^{d})$ iff:
\[
\begin{array}{l}
a: \cS(\rr^{d})\to \cS(\rr^{d}),\\[2mm]
{\rm ad}_{x}^{\alpha}{\rm ad}_{D}^{\beta}a\in B(H^{m- |\alpha|}(\rr^{d}), L^{2}(\rr^{d})), \ \forall\ \alpha, \beta\in \nn^{d},
\end{array}
\]
where ${\rm ad}_{x_{i}}b= [x_{i}, b]$, ${\rm ad}_{D_{j}}b= [D_{j}, b]$, and ${\rm ad}_{x}^{\alpha}= {\rm ad}_{x_{1}}^{\alpha_{1}}\cdots {\rm ad}_{x_{d}}^{\alpha_{d}}$ and similarly for ${\rm ad}_{D}^{\beta}$.

From (2) one can deduce (3) by expressing $f(a)$ for $f\in S^{p}(\rr)$ using the resolvent $(z- a)^{-1}$ and an almost analytic extension of $f$, see e.g. \cite{HS, D}.  Statement (4) follows from the parametrix construction of $(z-a)^{-1}$, which has $(z- \sigma_{\rm pr}(a))^{-1}$ as principal symbol. Statement (5) can be proved similarly. \qed

%

We conclude this subsection by stating a useful lemma, which follows from symbolic calculus.

\begin{lemma}\label{idiotic}
 Let $a\in \Psi^{p}(\rr^{d})$, $p\in \rr$,  $f,g\in \cinf(\rr^{d})$ with $\nabla f, \nabla g\in \coinf(\rr^{d})$ and $f\equiv 0$ near $\supp\,g$. Then  \[
f(x)a g(x)\in \Psi^{-\infty}(\rr^{d}).
\]
In particular $f(x)a g(x)$ maps $\cH'(\rr^{d})$ into $\cH(\rr^{d})$.
\end{lemma}

\subsection{Propagators}\label{sec1.5}

Let us fix a  map $\epsilon(t)=\epsilon_{1}(t)+ \epsilon_{0}(t)$, where $\epsilon_{i}(t)\in C^{\infty}(\rr, \Psi^{i}(\rr^{d}))$ for $i=0,1$. Assume moreover that 
$\epsilon_{1}(t)$  is elliptic in $\Psi^{1}(\rr^{d})$ and symmetric on $\cS(\rr^{d})$. It follows by Prop. \ref{1.1} that  $\epsilon_{1}(t)$ 
is selfadjoint with domain $H^{1}(\rr^{d})$, hence $\epsilon(t)$ with domain $H^{1}(\rr^{d})$ is closed, with non empty resolvent set.

We denote by $\Texp(\int_{s}^{t}\i\epsilon(\sigma)d\sigma)$ the associated  propagator defined by:

\[
\left\{
\begin{array}{rl}
&\frac{\p}{\p t}\Texp(\int_{s}^{t}\i\epsilon(\sigma)d\sigma)= \i  \epsilon(t)\Texp(\int_{s}^{t}\i\epsilon(\sigma)d\sigma),\\[2mm]
&\frac{\p}{\p s}\Texp(\int_{s}^{t}\i\epsilon(\sigma)d\sigma)= -\i  \Texp(\int_{s}^{t}\i\epsilon(\sigma)d\sigma)\epsilon(s),\\[2mm]
&\Texp(\int_{s}^{s}\i\epsilon(\sigma)d\sigma)=\one.
\end{array}
\right.
\]
Note that   the propagator $\Texp(\int_{s}^{t}\i\epsilon_{1}(\sigma)d\sigma)$ exists  and is unitary by e.g. \cite[Thm. X.70]{RS2}. Since $\epsilon(t)- \epsilon_{1}(t)$ is locally uniformly bounded in $B(L^{2}(\rr^{d}))$, one easily deduces  the existence of $\Texp(\int_{s}^{t}\i\epsilon(\sigma)d\sigma)$, which is strongly continuous in $(t,s)$ with values in $B(L^{2}(\rr^{d}))$.

\begin{definition}\label{def1}
 Assume in addition that $\epsilon(t)\in \Psi^{1}_{\rm ph}(\rr^{d})$.  Then we denote by $\Phi_{\epsilon}(t,s): \cooo{\rr^{d}}\to \cooo{\rr^{d}}$ the symplectic flow associated to  the time-dependent Hamiltonian $ -\sigma_{\rm pr}(\epsilon)(t, x, k)$. 
 
  Equivalently $ \Phi_{\epsilon}(t,s)$  is the restriction to the variables $(x, k)$ of the symplectic flow on $\cooo{\rr^{1+d}}$ associated to the Hamiltonian $\tau- \sigma_{\rm pr}(\epsilon)(t, x, k)$.
\end{definition}
Clearly $\Phi_{\epsilon}(t,s)$ is  an homogeneous map of degree $0$.

The following classical result is known as {\em Egorov's theorem}, see for instance \cite[Sec. 0.9]{taylor} for the proof. 
\begin{proposition}\label{1.2}
\ben
\item  $\Texp({\textstyle\int_{s}^{t}}\i\epsilon(\sigma)d\sigma)$ is bounded on $\cH(\rr^{d})$ hence on $\cH'(\rr^{d})$ by duality.

\item  Let $a\in \Psi^{m}(\rr^{d})$. Then 
 \[
a(t,s):=  \Texp({\textstyle\int_{s}^{t}}\i\epsilon(\sigma)d\sigma)a\Texp({\textstyle\int_{t}^{s}}\i\epsilon(\sigma)d\sigma) 
\]
belongs to $C^{\infty}(\rr^{2}, \Psi^{m}(\rr^{d}))$.   Moreover if $\epsilon(t)\in C^{\infty}(\rr, \Psi^{1}_{\rm ph}(\rr^{d}))$ and $a\in \Psi^{m}_{\rm ph}(\rr^{d})$ then $a(t,s)\in C^{\infty}(\rr^{2}, \Psi_{\rm ph}^{m}(\rr^{d}))$  
 with
 \[
\sigma_{\rm pr}(a)(t,s)= \sigma_{\rm pr}(a)\circ \Phi_{\epsilon}(s,t).
\] 
\een
\end{proposition}
From Proposition \ref{1.2} and Lemma \ref{1.0} we obtain the following result (the steps of the proof are explained in \cite[Sec. 0.10]{taylor}).
\begin{proposition}\label{1.2b}
For $u\in \cH'(\rr^{d})$ one has: 
\[
\WF(\Texp({\textstyle\int_{s}^{t}}\i\epsilon(\sigma)d\sigma)u)= \Phi_{\epsilon}(t,s)\WF(u),
\]
hence
\[
\WF(\Texp({\textstyle\int_{s}^{t}}\i\epsilon(\sigma)d\sigma))'= \{(x, k, x', k')  : \ (x, k)= \Phi_{\epsilon}(t,s)(x', k')\}.
\]
\end{proposition}
\begin{lemma}\label{buda}
 Let $\epsilon(t)\in \cinf(\rr, \Psi^{1}(\rr^{d}))$ as above,  $s_{-\infty}(t,s)\in C^{\infty}(\rr^{2}, \Psi^{-\infty}(\rr^{d}))$. Then
 \[
\Texp({\textstyle\int_{s}^{t}}\i\epsilon(\sigma)d\sigma) s_{-\infty}(t,s)\in C^{\infty}(\rr^{2}, \Psi^{-\infty}(\rr^{d})).
\]
\end{lemma}
\proof The proof will be given in Subsect. \ref{budaproof}. \qed

\section{Concrete Klein-Gordon equations}\label{sec2}\init

\subsection{Model Klein-Gordon equation on $\rr^{1+d}$}\label{sec2.0b}
In this subsection we describe the model Klein-Gordon equation that will be considered in the sequel.  We take $M= \rr^{1+d}$, $x= (t, \rx)\in \rr^{1+d}$
 and fix a second order differential operator
\beq\label{e0.3b}
a(t, \rx, D_{\rx})= -\sum_{j,k=1}^{d}\p_{\rx^{j}}a^{jk}(x)\p_{\rx^{k}}+ \sum_{j=1}^{d} b^{j}(x)\p_{\rx^{j}}- \p_{\rx^{j}}\overline{b}^{j}(x)+ m(x),
\eeq
where 
\begin{equation}
\label{e0.3}
\begin{array}{l}
a^{jk}, b^{j}, m \in C^{\infty}(\rr, C^{\infty}_{\rm bd}(\rr^{d})), \   m(x)\in \rr,\\[2mm]
[a^{jk}](x)\geq  c(t)\one\hbox{ uniformly on }\rr^{1+d}, \ c(t)>0. 
\end{array}
\end{equation}

 We introduce the model Klein-Gordon operator
 \[
P(x, D_{x})= \p_{t}^{2}+ a(t, \rx, D_{\rx}), 
\]
which is formally selfadjoint for the scalar product $(u_{1}| u_{2})= \int_{\rr^{1+d}} \overline{u}_{1}u_{2}dx$.

We will consider the Cauchy problem:
\begin{equation}
\label{e3.1}
\left\{
\begin{array}{rl}
&\p_{t}^{2}\phi(t)+a(t, \rx, D_{\rx})\phi(t)=0,\\[2mm]
&\phi(s)=f_{0},\\[2mm]
&\i^{-1}\p_{t} \phi(s)=f_{1},
\end{array}
\right.
\end{equation}
for $f= (f_{0}, f_{1})\in \cH(\rr^{d})\otimes \cc^{2}$. By the well-known method of energy estimates, one obtains the existence and uniqueness of a solution $\phi\in \cinf(\rr, \cH(\rr^{d}))$. Similarly if $f\in \cH'(\rr^{d})\otimes \cc^{2}$, there exists a unique solution $\phi\in \cinf(\rr, \cH'(\rr^{d}))$.

\subsection{Reduction to the model case}\label{ss:modelcase}
Consider a globally hyperbolic space-time $(M, g)$ with a Cauchy hypersurface diffeomorphic to $\rr^{d}$. 
This implies that we can assume that $M= \rr_{t}\times \rr^{d}$ and
\beq\label{metric}
g= -c (x)dt^{2}+ h_{jk}(x)d\rx^{j}d\rx^{k},
\eeq
where $x= (t, \rx)$, $c(x)>0$ is a smooth function and $h_{jk}(x)d\rx^{j}d\rx^{k}$ is a smooth Riemannian metric on $\rr^{d}$.  

In this subsection we explain how to reduce the  Klein-Gordon operator (\ref{e0.00}) to the model case considered in Subsect. \ref{sec2.0b}.

Writing $A_{\mu}(x)= (V(x), \rA_{j}(x))$, we have:
\[\begin{array}{rl}
&P(x, D_{x})\\[2mm]
=&\ c^{-\12}|h|^{-\12}(\p_{t}+ \i V) c^{-\12}|h|^{\12}(\p_{t}+\i V)\\[2mm]
&- c^{-\12}|h|^{-\12}(\p_{j}+\i A_{j})c^{\12}| h|^{\12}h^{jk}(\p_{k}+\i A_{k})+ \rho,
\end{array}
\]
where $|h|= \det[h_{jk}]$, $ [h^{jk}]= [h_{jk}]^{-1}$.  

We choose the Cauchy hypersurface $S=\{0\}\times \rr^{d}$ so that
\beq\label{e2.0}
\bar\phi_{1}\sigma \phi_{2}= \int_{S}\left(\overline{(\p_{t}+\i V)\phi_{1}}\phi_{2}- \overline{\phi_{1}}(\p_{t}+ \i V)\phi_{2}\right)c^{-\12}| h|^{\12}d \rx.
\eeq
Set:
\[
\begin{array}{rl}
F(t, \rx)=&  \int_{0}^{t}V(s, \rx)ds, \ \tilde{\rA}=\rA+ \nabla F,\  \tilde{\rho}= c \rho- c^{\41}|h|^{-\41}\p_{t}^{2}(c^{-\41}|h|^{\41}),\\[2mm]
a(t, \rx, D_{\rx})= &c^{\41}|h|^{-\41}(\p_{j}+\i \tilde{\rA}_{j})c^{\12}|h|^{\12}h^{jk}(\p_{k}+\i \tilde{\rA}_{k})c^{\41}|h|^{-\41}+ \tilde{\rho}.
\end{array}
\]
\begin{lemma}\label{2.1}
\ben
\item Let $b= c^{-\frac{1}{4}}|h|^{\frac{1}{4}}$. Then:
\[
P(x, D_{x})= c^{-\12}|h|^{-\12} b\e^{-\i F}\left(\p_{t}^{2}+ a(t, \rx, D_{\rx})\right)b\e^{\i F},
\]
hence
\[
 E= b^{-1}\e^{- \i F}\tilde{E}c^{\12}|h|^{\12}b^{-1}\e^{\i F},
\]
where $\tilde{E}$ is the Pauli-Jordan function for $\tilde{P}(x, D_{x}):= \p_{t}^{2}+ a(t, \rx, D_{\rx})$.
\item The map:
\[
\phi\mapsto \tilde{\phi}:=c^{-\41}|h|^{\41}\e^{-\i F}\phi, 
\]
is symplectic from  $({\rm Sol}_{\rm sc}(P), \sigma)$  to $({\rm Sol}_{\rm sc}(\tilde{P}), \tilde{\sigma})$ for
\[
\overline{\tilde{\phi}_{1}}\tilde{\sigma}\tilde{\phi}_{2}=  \int_{S}\overline{\p_{t}\tilde{\phi}_{1}}\tilde{\phi}_{2}-  \overline{\tilde{\phi}_{1}}\p_{t}\tilde{\phi}_{2}d\rx.
\]
\item Assume the following hypotheses:
\[
(H) \begin{array}{rl}
&\forall \ I\subset \rr\hbox{ compact interval }\exists\  C>0\hbox{ such that }\\[2mm]
&C\leq c(x), \ C\one\leq [h_{jk}(x)],\hbox{uniformly for }x\in I\times \rr^{d},\\[2mm]
&  h_{jk}(x), \ c(x), \ \rho(x), A_{\mu}(x)\in C^{\infty}(\rr, C^{\infty}_{\rm bd}(\rr^{d})).
\end{array}
\]
Then the operator $a(t, \rx, D_{\rx})$ is of the form (\ref{e0.3b}) and the conditions (\ref{e0.3}) are satisfied.
\een
\end{lemma}
\proof
Set
\[
b^{2}= c^{-\12}|h|^{\12}, \ a= (\p_{j}+\i A_{j})c^{\12}| h|^{\12}h^{jk}(\p_{k}+\i A_{k})+ c^{\12}|h|^{\12}\rho.
\] 
Then 
\[
\begin{array}{rl}
P(x, D_{x})= &c^{-\12}|h|^{-\12}((\p_{t}+ \i V)b^{2}(\p_{t}+ \i V)+a),\\[2mm]
=&c^{-\12}|h|^{-\12}b((\p_{t}+ \i V)^{2}+ \tilde{a})b,
\end{array}
\]
for $ \tilde{a}=  b^{-1} ab^{-1}- b^{-1}(\p_{t}^{2}b)$.    Since $\p_{t}+ \i V= \e^{- \i F}\p_{t}\e^{\i F}$ we finally get:
\[
P(x, D_{x})= c^{- \12}|h|^{-\12} b\e^{- \i F}\left(\p_{t}^{2}+ a(t, \rx, D_{\rx})\right)\e^{\i F}b,
\] 
  This proves (1). (2) and (3) are  left to the reader. \qed

By Lemma \ref{2.1}, the task of constructing Hadamard states for $P(x,D)$ can be reduced to constructing Hadamard states for the model Klein--Gordon equation $\tilde P(x,D)$. Indeed, suppose we have a Hadamard state with two-point function $\tilde{\Lambda}$ and charge $\i \tilde E$. Then 
\[
\Lambda:= b^{-1}\e^{- \i F}\tilde{\Lambda}c^{\12}|h|^{\12}b^{-1}\e^{\i F}
\]
defines the two-point function of a Hadamard state with charge $\i  E$, the wave front set being preserved by multiplication by smooth densities.

\section{The model Klein-Gordon equation}\label{sec3}\init
In this section we consider the model Klein-Gordon operator $P(x, D_{x})= \p_{t}^{2}+ a(t, \rx, D_{\rx})$ introduced in Subsect. \ref{sec2.0b}.  The associated symplectic form is:
\[
\overline{\phi}_{1}\sigma\phi_{2}=  \int_{{t}\times \rr^{d}}\overline{\p_{t}\phi_{1}}\phi_{2}-  \overline{\phi_{1}}\p_{t}\phi_{2}d\rx.
\]
\subsection{Notation}
If $a\geq 0$ is a selfadjoint operator on a Hilbert space $\mathfrak{h}$ we write $a>0$ if $\Ker\, a= \{0\}$. Then $a^{-1}$ with domain ${\rm Ran}\,a$ is selfadjoint.

If $a,b\geq 0$ are two selfadjoint operators  on a Hilbert space $\mathfrak{h}$ then we write $a\leq b$ if $\Dom b^{\12}\subset \Dom a^{\12}$ and $(u| au)\leq (u| b u)$ for $u\in \Dom b^{\12}$. 

The Kato-Heinz theorem  says that if $0\leq a\leq b$ then $0\leq a^{s}\leq b^{s}$ for all $0\leq s \leq 1$ and if 
$0<a\leq b$ then $0\leq b^{-1}\leq a^{-1}$.

We write $a\sim b$ if $c^{-1}a\leq b\leq c a$ for some $c>0$.
\subsection{Bicharacteristic curves}
We denote
$a_{2}(t, \rx, k)= k_{i}a^{ij}(t, \rx)k_{j}$ the principal symbol of $a(t, \rx, D_{\rx})$ and by $p(x,\xi)= - \tau^{2}+ a_{2}(t, \rx, k)$ the principal symbol of $P(x, D_{x})= \p_{t}^{2}+ a(t,\rx, D_{\rx})$. 


We set:
\[
\epsilon_{1}(t, \rx, k)= (k_{i}a^{ij}(t, \rx)k_{j})^{\12}.
\]
As in Def. \ref{def1} we denote by $\Phi_{\pm}(t,s): \cooo{\rr^{d}}\to \cooo{\rr^{d}}$ the restrictions to  the variables 
$(\rx, k)$ of the symplectic flows on $\cooo{\rr^{1+d}}$ associated to the hamiltonians $\tau\mp \epsilon_{1}(t, \rx, k)$.

The following lemma is immediate:
\begin{lemma}\label{2.3}
Let $X_{i}\in \cN_{\pm}$, $i=1,2$ with $X_{1}\sim X_{2}$. Then
 \[
X_{i}= (t_{i}, \rx_{i}, \pm\epsilon_{1}(t_{i}, \rx_{i}, k_{i}), k_{i})\hbox{ with } 
(\rx_{2}, k_{2})= \Phi_{\pm}(t_{2}, t_{1})(\rx_{1}, k_{1}).
\]
\end{lemma}


\subsection{Parametrix for the Cauchy problem}\label{sec3.2}
In this subsection we outline the well-known construction of a parametrix for the Cauchy problem (\ref{e3.1}).
The construction is well-known and belongs to the folklore of microlocal analysis. Usually it is done using  Fourier integral operators.
Our construction relies more on Hilbert space methods.

We start by an auxiliary lemma.
\begin{lemma}\label{2.2}
Assume (\ref{e0.3}) and let $a(t, \rx, D_{\rx})$ be given by (\ref{e0.3b}). Then there exists  smooth maps 
\[
\begin{array}{rl}
\rr\ni t&\mapsto\epsilon(t)= \epsilon(t, \rx, k)\in S^{1}_{\rm ph}(\rr^{2d}),\\[2mm]
\rr\ni t&\mapsto r_{-\infty}(t)= r_{-\infty}(t, \rx, k)\in S^{-\infty}(\rr^{2d}),
\end{array}
\]
such that:
\ben
\item $\epsilon(t, \rx, k)$ is real-valued, $\epsilon(t, \rx, k)\in C^{\infty}(\rr, S^{1}_{\rm ph}(\rr^{2d}))$ with principal symbol
\[
\epsilon_{1}(t, \rx, k)= (k_{i}a^{ij}(t, \rx)k_{j})^{\12},
\]
\item $\epsilon^{\rm w}(t, \rx, D_{\rx})\geq c(t)(D_{x}^{2}+\one)^{\12}$ for $c(t)>0$,
\item   $a(t, \rx, D_{\rx})=\epsilon^{\rm w}(t, \rx, D_{\rx})^{2}- r_{-\infty}^{\rm w}(t, \rx, D_{\rx})$.
\een
Moreover $\epsilon(t)$ and $r_{-\infty}(t)$ are unique modulo $\cinf(\rr, \Psi^{-\infty}(\rr^{d}))$.
\end{lemma}
\proof  The proof will be given in Subsect. \ref{sec-app1.1bis}. \qed 

The following theorem is the main result of this section. It will be used later on to characterize and construct examples of quasi-free Hadamard states.

\begin{theoreme}\label{3.1}
There exist  operators $b(t)\in \cinf(\rr, \Psi^{1}_{\rm ph}(\rr^{d}))$, $r(t)\in \cinf(\rr, \Psi^{-1}_{\rm ph}(\rr^{d}))$  with
\[
\begin{array}{rl}
(i)& b(t)=\epsilon(t)+ (2\epsilon(t))^{-1} \i \p_{t}\epsilon(t)\hbox{ mod }\cinf(\rr, \Psi^{-1}(\rr^{d})),\\[2mm]
(ii)&r(t)=b^{*}(t)^{(-1)}\hbox{ mod }\cinf(\rr, \Psi^{-\infty}(\rr^{d})),\\[2mm]
(iii)&r(t)= \epsilon(t)^{-1}+ \cinf(\rr, \Psi^{-2}(\rr^{d})),\\[2mm]
(iv)&r(t)+ r(t)^{*}\sim \epsilon(t)^{-1},
\end{array}
\]
such that if
\[
\begin{array}{rl}
u_{+}(t,s):=& \Texp(\i\int_{s}^{t}b(\sigma)d\sigma),\ u_{-}(t,s):= \Texp(-\i \int_{s}^{t}b^{*}(\sigma)d \sigma)\\[2mm]
d_{+}(t):=&(\one +b^{*}(t)^{(-1)}b(t))^{(-1)},\ d_{-}(t):= d_{+}(t)^{*},\\[2mm]
r_{+}(t):=& r(t), \ r_{-}(t):= r^{*}(t), 
\end{array}
\]
the following properties hold:
\ben
\item set for $f\in \cH'(\rr^{d})\otimes \cc^{2}$:
\[
\begin{array}{rl}
U(t,s)f= &u_{+}(t,s)d_{+}(s)\left(f_{0}+ r_{+}(s)f_{1}\right)+ u_{-}(t,s)d_{-}(s)\left(f_{0}- r_{-}(s)f_{1}\right),\\[2mm]
=:& U_{+}(t,s)f+ U_{-}(t,s)f.
\end{array}
\]
then 
\beq\label{e.buda5}
\left\{\begin{array}{l}
(\p_{t}^{2}+ a(t, \rx, D_{\rx}))U(t,s)f= s_{-\infty}(t,s)f,\\[2mm]
U(s,s)f= f_{0}+ r_{-\infty,0}(s)f,\\[2mm]
 \i^{-1}\p_{t}U(t,s)f|_{t=s}= f_{1}+ r_{-\infty, 1}(s)f,
\end{array}\right.
\eeq
for $s_{-\infty}(t,s)\in \cinf(\rr^{2}, \Psi^{-\infty}(\rr^{d}))\otimes \cc^{2}$, 
$r_{-\infty, i}(s)\in \cinf(\rr, \Psi^{-\infty}(\rr^{d}))\otimes \cc^{2}$.

\item  let $\phi(t)$ be the unique solution of (\ref{e3.1}) for $f\in \cH'(\rr^{d})\otimes \cc^{2}$. Then:
\beq\label{e3.02}
\phi(t)=U(t,s)f { \ \rm mod \ } C^{\infty}(\rr, \cH(\rr^{d})).
\eeq
\een
\end{theoreme}
\proof  the proof will be given in Subsect. \ref{sec-app1.2}. \qed

To simplify notation, in the rest of the paper, we will fix $s=0$, and set:
\[
\begin{array}{rl}
&b:=b(0), \ u_{\pm}(t):=u_{\pm}(t,0), \ U(t):= U(t,0),\\[2mm]
&d:=d(0), \ r:= r(0), \ \epsilon:= \epsilon(0).
\end{array}
\]

The constructions of Hadamard states in Sect. \ref{sec4} will a priori depend on the choice of an operator $r$. To study this dependence it is convenient to introduce the following definition.

\begin{definition}\label{defo}
 We denote by $\cM$ the set:
 \[
\cM:= \{r\in \Psi^{-1}_{\rm ph}(\rr^{d})\ : \ r= b^{*(-1)}+ \Psi^{-\infty}(\rr^{d}), \   r+ r^{*}\sim  \epsilon^{-1}\}.
\]
\end{definition}
\begin{remark}
Note that the operator $b$ in Thm. \ref{3.1} is unique modulo $\Psi^{-\infty}$.   Thm. \ref{3.1} implies  that $\cM$ is not empty. Since two elements of $\cM$ are equal modulo $\Psi^{-\infty}$, the conclusions of Thm. \ref{3.1} are valid for any $r\in \cM$.
\end{remark}
The following corollary is a consequence of (\ref{e3.02}), Prop. \ref{1.2b} and Lemma \ref{2.3}.
\begin{corollary}\label{3.1b}
If $\phi(t)$ is the unique solution of (\ref{e3.1}) for $f\in \cH'(\rr^{d})\otimes \cc^{2}$, one has
\[
\phi(t)= U_{+}(t)f+ U_{-}(t)f { \ \rm mod \ } C^{\infty}(\rr, \cH(\rr^{d})),
\]
and\[\begin{array}{rl}
& \WF(U_{\pm}(t)f)\\[2mm]
=&\{(x_{2}, \xi_{2})  : \ \exists \ (\rx_{1}, k_{1})\in \WF (f_{0}\pm r_{\pm}f_{1})\hbox{ with }(s, \rx_{1}, \pm \epsilon(\rx_{1}, k_{1}), k_{1})\sim (x_{2}, \xi_{2})\}.
\end{array}
\]
In particular \[
\WF(U_{\pm}(t)f)\subset \cN_{\pm}.
\]
\end{corollary}
\subsection{Symplectic properties of the spaces of positive/negative wavefront set solutions}\label{sec3.3}
We now investigate the properties, with respect to the symplectic form $\sigma$, of the spaces of solutions of the Klein-Gordon equation having wavefront set included in  the positive/negative energy surfaces $\cN_{\pm}$. 

Of course we cannot work with  solutions in $\cE(M)$, since their wavefront set is empty, nor with solutions in $\cD'(M)$, since the symplectic form $\sigma$ is not defined on arbitrary distributional solutions. A natural space of solutions is the space of {\em finite energy solutions} defined as follows:
\[
{\rm Sol}_{E}(P):= \{\phi\in C^{0}(\rr, H^{1}(\rr^{d}))\cap C^{1}(\rr, L^{2}(\rr^{d}))\ :\ P(t, \rx, D_{\rx})\phi=0\}.
\]
It is well-known that  $\phi\in{\rm Sol}_{E}(P)$ iff $f=(\phi(0), \i^{-1}\p_{t}\phi(0))\in H^{1}(\rr^{d})\oplus L^{2}(\rr^{d})$.
 Moreover the symplectic form $\sigma$ is well defined in ${\rm Sol}_{E}(P)$.

Recall that for  $r\in \cM$ one sets $r_{+}= r, \ r_{-}= r^{*}$. We define now
\[
C^{\pm}(r):= \{f\in H^{1}(\rr^{d})\oplus L^{2}(\rr^{d})\ : \ f_{0}\mp r_{\mp}f_{1}=0\},
\]
and
\[
{\rm Sol}_{E}^{\pm}(P,r):= \{\phi\in {\rm Sol}_{E}(P)\ : \ (\phi(0); \i^{-1}\p_{t}\phi(0))\in C^{\pm}(r)\}.
\]
We call ${\rm Sol}_{E}^{\pm}(P,r)$ the space of {\em  positive/negative wavefront set solutions}.
\begin{theoreme}\label{posit}
 Let $r\in\cM$. Then the following properties hold:
 \ben
 \item ${\rm Sol}_{E}(P)= {\rm Sol}^{+}_{E}(P,r)\oplus{\rm Sol}^{-}_{E}(P,r)$,
 \item $\phi\in {\rm Sol}^{\pm}_{E}(P,r)\ \Rightarrow \ \WF(\phi)\subset \cN_{\pm}$,
 \item $\pm\i \sigma= \pm q>0$ on ${\rm Sol}^{\pm}_{E}(P,r)$,
 \item the spaces $ {\rm Sol}^{\pm}_{E}(P,r)$ are symplectically orthogonal. 
   \een
\end{theoreme}
\begin{remark}
 We can interpret Thm. \ref{posit} as follows:  the space of finite energy solutions decomposes as the direct sum of the spaces of positive resp. negative wavefront set solutions. The charge $q$ is positive, resp. negative on these spaces. Moreover these two spaces are symplectically orthogonal. 
 This is the exact analogue of the well-known situation in the static case,
where $a(t, \rx, D_{\rx})$ does not depend on $t$ (cf. Subsect. \ref{for-record}). 
\end{remark}
For the proof of Thm. \ref{posit}, we will use the following lemma.
\begin{lemma}\label{4.1}
Let $r\in \cM$. Then:
 \ben
 \item $r+ r^{*}: L^{2}(\rr^{d})\to H^{1}(\rr^{d})$ is invertible  and 
 \[
 (r+ r^{*})^{-\12}= \frac{1}{\sqrt{2}} \epsilon^{\12}+ \Psi_{\rm ph}^{0}(\rr^{d}).
 \]
 \item The operator 
 \[
T(r) := (r+r^{*})^{-\12}\mat{\one}{r}{\one}{-r^{*}}: \  \cH\shit\to  \cH\shit,
\]
is bounded with bounded inverse:
\[
T(r)^{-1}= \mat { r^{*}}{r}{\one}{-\one}(r+r^{*})^{-\12}.
\]
\item $T(r): H^{1}(\rr^{d})\oplus L^{2}(\rr^{d})\to H^{\12}(\rr^{d})\otimes \cc^{2}$ is bounded with inverse $T(r)^{-1}$.
\item one has:
\[
\tilde{q}:= (T(r)^{-1})^{*}\circ q\circ T(r)^{-1}= \mat{\one}{0}{0}{-\one}.
\]
 \een
\end{lemma}
\proof 
From Thm.  \ref{3.1}  we obtain that $r+ r^{*}\sim  \epsilon^{-1}$, which implies that $r+r^{*}$ is bijective from $L^{2}(\rr^{d})$ to $H^{1}(\rr^{d})$. Moreover $(r+r^{*})^{-1}= \half \epsilon+ \Psi^{0}$ and  $(r+r^{*})^{-1}\sim  \epsilon$. By Prop. \ref{1.1} we obtain (1). Statements (2), (3), (4) follow from (1). \qed

\medskip

{\bf Proof of Thm. \ref{posit}.}
For $\tilde{f}\in H^{\12}(\rr^{d})\otimes \cc^{2}$ we set $\tilde{f}= (\tilde{f}_{+}, \tilde{f}_{-})$.  Since $r_{+}= r$, $r_{-}= r^{*}$ we obtain that by Lemma \ref{4.1} $f\in C^{\pm}$ iff $(Tf)_{\mp}=0$.  The theorem follows then from Lemma \ref{4.1} (4).   \qed
\section{Construction of Hadamard states}\label{sec4}\init
\subsection{Microlocal spectrum condition}\label{sec4.2}

In this subsection we discuss conditions under which a covariance $c$ on $\cD\shit$ leads by(\ref{defino}) to a covariance $C$ on $\cD(M)$ satisfying the microlocal spectrum condition in Def. \ref{def2}.

Recall that  $c:E_{1}\to E_{2}$ means that $c$ is linear continuous between the two topological vector spaces $E_{1}$ and $E_{2}$. 

We consider the model Klein-Gordon equation:
\[
\p_{t}^{2}\phi + a(t, \rx, D_{\rx})\phi=0,
\]
introduced in Subsect. \ref{sec2.0b}.

Let $c$  be a  bounded hermitian form on  $\cD(\rr^{d})\otimes \cc^{2}$. We identify it with the operator:
\beq\label{e4.001}
c= \mat{c_{00}}{c_{01}}{c_{10}}{c_{11}}: \ \cD(\rr^{d})\otimes \cc^{2}\to \cD'(\rr^{d})\otimes \cc^{2},
\eeq
and associate to it the  bounded hermitian form  $C$ on $\cD(\rr^{1+d})$ given by:
\beq\label{e4.01}
(u|Cv)= (\rho\circ E u|c \rho\circ Ev), \ C: \cD(\rr^{1+d})\to \cD'(\rr^{1+d}).
\eeq
We still denote  by $C\in \cD'(\rr^{1+d}\times \rr^{1+d})$ the distribution kernel of $C$ given by:
\[
(u|Cv)= \int C(x, y)\overline{u}(x)v(y)dxdy.
\] 

We fix now an operator $r\in \cM$ (see Def. \ref{defo}). The map $T(r)$ in Lemma \ref{4.1} will be denoted by $T$ for simplicity.

We would like to define:
\beq\label{e4.05}
\tilde{c}:= (T^{-1 })^{*}\circ c \circ T^{-1}=: \mat{\tilde{c}_{++}}{\tilde{c}_{+-}}{\tilde{c}_{-+}}{\tilde{c}_{--}}.
\eeq
Since $T$, $T^{-1}: \cH\shit\to \cH'\shit$, a natural requirement is that
\[
c: \cH\shit\to \cH'\shit,
\]
which implies that $c: \cD\shit\to \cD'\shit$.  In the next theorem we will need to impose stronger conditions on $c$.

We can now state the main result of this subsection.  Recall that the notations $ _{M}\!\Gamma$ and $\Gamma_{M}$  for a conic set $\Gamma$ are defined in Subsect. \ref{sec-1}. 
\begin{theoreme}\label{4.2}
Assume that 
\[
 \begin{array}{l}
(1a)\ c: \cH\shit\to \cH\shit, \\[2mm]
(1b)\ \ _{\rr^{d}}\!\WF(c)'= \WF(c)'_{\rr^{d}}= \emptyset. 
\end{array}
\] 
Let $C$ be defined by (\ref{e4.01})  and 
\[
 \Lambda_{+}:= C, \Lambda_{-}:= C- \i E.
\]
Then $\Lambda_{\pm}$ satisfies the microlocal spectrum condition iff:
\[
(2)\ \WF(\tilde{c}_{--})'= \WF(\tilde{c}_{+-})'= \WF(\tilde{c}_{-+})'=  \WF(\one -\tilde{c}_{++})'= \emptyset.
\]
\end{theoreme}
\begin{remark}\label{remaro}
 Note that condition (2) implies that condition (1b) is satisfied by $\tilde{c}$.
 Using that $T$, $T^{-1}$ are pseudo-differential  operators, it is easy to see that condition (1b) is then also satisfied by $c$. Therefore if conditions (1a), (2) are satisfied, $\Lambda_{\pm}$ satisfies $(\mu{\rm sc})$.
\end{remark}
\begin{remark}
Note  that we strengthen the assumption on the sesquilinear form $c$, since we require in (1a) that $c: \cH\shit\to \cH\shit$ instead of  $c: \cH\shit\to \cH'\shit$ as before.  In fact since the Cauchy surface is not compact, some care is needed with the composition  of operators. 

Condition (1b) is satisfied for example if $\WF(c)'\subset \Gamma$, where $\Gamma$ is the graph of a conic, bijective map on $T^{*}\rr^{d}$.
This is the case if the entries of $c$ are pseudo-differential  or even Fourier integral operators. 
\end{remark}
\proof

We set  \[
\tilde{\rho}= T\circ \rho=: \tilde\rho_{+}\oplus \tilde\rho_{-}, 
\]so that:
\beq\label{e4.04}
C= (\tilde{\rho}\circ E)^{*}\circ \tilde{c}\circ (\tilde{\rho}\circ E)=\sum_{\alpha, \beta\in\{+,-\}}C_{\alpha \beta},
\eeq
for
\beq\label{e4.04b}
 \ C_{\alpha \beta}:= (\tilde\rho_{\alpha}\circ E)^{*}\circ \tilde{c}_{\alpha\beta}\circ (\tilde\rho_{\beta}\circ E).
\eeq
Let us first check that we can perform the various compositions in (\ref{e4.04}).

Because of the well-known support properties of $E_{\pm}$  we have:
\beq\label{shiti}
\begin{array}{rl}
\rho\circ E:&\begin{array}{l}
\cD(M)\to \cD(\rr^{d})\otimes \cc^{2},\\[2mm]
\cE'(M)\to \cE'\shit,
\end{array}\\[4mm]
(\rho\circ E)^{*}:&\begin{array}{l}
\cD'(\rr^{d})\otimes \cc^{2}\to \cD'(M),\\[2mm]
\cE\shit\to \cE(M).
\end{array}
\end{array}
\eeq
Note also that  $T: \cD(\rr^{d})\otimes \cc^{2}\to \cH(\rr^{d})\otimes \cc^{2}$ and $T: \cE'\shit\to \cH'\shit$. We obtain that 
\beq\label{shito}
\begin{array}{rl}
\tilde\rho\circ E:&\begin{array}{l}
\cD(M)\to \cH(\rr^{d})\otimes \cc^{2},\\[2mm]
\cE'(M)\to \cH'\shit,
\end{array}\\[4mm]
(\tilde\rho\circ E)^{*}:&\begin{array}{l}
\cH'(\rr^{d})\otimes \cc^{2}\to \cD'(M),\\[2mm]
\cH\shit\to \cE(M).
\end{array}
\end{array}
\eeq

Since we assumed that $c:\cH(\rr^{d})\otimes \cc^{2}\to \cH'(\rr^{d})\otimes \cc^{2}$, we have $\tilde{c}: \cH(\rr^{d})\otimes \cc^{2}\to \cH'(\rr^{d})\otimes \cc^{2}$, using that $T$, $T^{-1}$ preserve $\cH(\rr^{d})\otimes \cc^{2}$ and $\cH'(\rr^{d})\otimes \cc^{2}$. Therefore we can perform the compositions in (\ref{e4.04}).

By Lemma \ref{lem:raff}, we have
\[
\WF(E)'=\{(X_{1}, X_{2}) : \ X_{1}\sim X_{2}, \ X_{1}, X_{2}\in \cN\},
\]
and using that the Cauchy surface $\{t=0\}$ is  non-characteristic for the Klein-Gordon equation, we have for $i=0,1$:
\[
\begin{array}{rl}
&\WF(\rho_{i}\circ E)'\\[2mm]
=& \{\left((\rx_{1}, k_{1}), (x_{2}, \xi_{2})\right)\in \coo{(\rr^{d}\times M)} : \ (0, \rx_{1}, - \epsilon_{1}(0, \rx_{1}, k_{1}), k_{1})\sim (x_{2}, \xi_{2})\} \\[2mm]
 \cup & \{\left((\rx_{1}, k_{1}), (x_{2}, \xi_{2})\right)\in \coo{(\rr^{d}\times M)} : \ (0, \rx_{1}, + \epsilon_{1}(0, \rx_{1}, k_{1}), k_{1})\sim (x_{2}, \xi_{2})\}.
\end{array}
\]
Then from Corollary \ref{3.1b}, we obtain that:
\begin{equation}
\label{e4.1}
\WF(\tilde\rho_{\pm}\circ E)'= \Gamma_{\pm},
\end{equation}
for
\[
\Gamma_{\pm}=\{\left((\rx_{1}, k_{1}), (x_{2}, \xi_{2})\right)\in \coo{(\rr^{d}\times M)} : \ (0, \rx_{1}, \pm \epsilon(0, \rx_{1}, k_{1}), k_{1})\sim (x_{2}, \xi_{2})\}.
\]
We also have
\beq\label{e4.1bis}
\WF((\tilde\rho_{\pm}\circ E)^{*})'= {\rm Exch}(\Gamma_{\pm}).
\eeq
We now want to apply the composition rule for the wave front set recalled in Subsect. \ref{sec-1} to the identity (\ref{e4.04b}), in order to bound $\WF(C_{\alpha\beta})'$.  It clearly suffices to bound $\WF(\chi C_{\alpha\beta}\chi)'$ for $\chi\in \coinf(M)$. 

\medskip

{\em Step 1.}
The first step is to reduce oneself  to the composition of properly supported kernels, modulo smoothing operators.  

Because of the support properties of the kernel of $E$, there exists $\psi\in \coinf(\rr^{d})$ such that (denoting again $\mat{\psi}{0}{0}{\psi}$ by $\psi$):
\[
 \rho\circ E \chi= \psi \rho\circ E \chi.
\]
Let us also fix $\psi_{1}\in \coinf(\rr^{d})$ with $\psi_{1}\equiv 1$ near $\supp \psi$.  By Lemma \ref{idiotic} we know that $ (1- \psi_{1})T\psi\in \Psi^{-\infty}(\rr^{d})$, hence
\[
\tilde\rho\circ E \chi= \psi_{1}\tilde\rho\circ E \chi+ R_{1}\rho\circ E \chi,
\]
for
\beq\label{shitu}
R_{1}:= (1- \psi_{1})T \psi: \cD'\shit\to \cH\shit.
\eeq
Taking adjoints we have:
\[
\begin{array}{l}
\chi(\tilde\rho\circ E)^{*}= (\tilde \rho \circ E \chi)^{*}= \chi(\tilde \rho\circ E)^{*}\psi_{1}+ \chi (\rho\circ E)^{*}R_{1}^{*},\\[2mm]
R_{1}^{*}: \cH'\shit \to \cD\shit.
\end{array}
\]
It follows that 
\[
\begin{array}{rl}
& \chi(\tilde \rho\circ E)^{*}\tilde{c}(\tilde \rho\circ E)\chi\\[2mm]
=&  \chi(\tilde \rho\circ E)^{*}\psi_{1}\tilde{c}\psi_{1}(\tilde \rho\circ E)\chi
+  \chi(\tilde \rho\circ E)^{*}\tilde{c}R_{1}(\rho\circ E)\chi+  \chi(\tilde \rho\circ E)^{*}R_{1}^{*}\tilde{c}\psi_{1}(\tilde \rho\circ E)\chi\\[2mm]
=:& \chi(\tilde \rho\circ E)^{*}\psi_{1}\tilde{c}\psi_{1}(\tilde \rho\circ E)\chi+ I_{1}+ I_{2}.
\end{array}
\]
We claim that 
\beq\label{sta}
I_{1}, \ I_{2}: \cD'(M)\to \cD(M).
\eeq
Note that from hypothesis (1a) and the fact that $T$, $T^{-1}$ are (matrices of) pseudo-differential  operators, we know that $\tilde{c}: \cH\shit\to \cH\shit$.

Using then (\ref{shiti}), (\ref{shito}), (\ref{shitu}) we have:
\[
\begin{array}{rl}
I_{1}:&\cD'(M):\mathop{\to}\limits^{(\rho\circ E)\chi} \cE'\shit\subset \cH'\shit\mathop{\to}\limits^{R_{1}} \cH\shit\\[2mm]
 &\mathop{\to}\limits^{\tilde{c}}\cH\shit\subset \cE\shit\mathop{\to}\limits^{\chi(\tilde{\rho}\circ E)^{*}} \cD(M),
\end{array}
\]
and:
\[
\begin{array}{rl}
I_{2}:&\cD'(M):\mathop{\to}\limits^{(\tilde\rho\circ E)\chi} \cH'\shit\mathop{\to}\limits^{\tilde{c}\psi_{1}} \cH'\shit\\[2mm]
 &\mathop{\to}\limits^{R_{1}^{*}}\cD\shit\subset \cE\shit\mathop{\to}\limits^{\chi(\rho\circ E)^{*}} \cD(M),
\end{array}
\]
which proves (\ref{sta}).  It follows that if $\psi_{2}\in \coinf(\rr^{d})$ with $\psi_{2}\equiv 1$ near $\supp \psi_{1}$ we have:
\[
\chi C_{\alpha\beta} \chi=   \chi (\tilde \rho_{\alpha}\circ E)^{*} \psi_{1}\circ \psi_{2}\tilde{c}\psi_{2}\circ \psi_{1}(\tilde{\rho}_{\beta}\circ E)\chi\hbox{ mod }C^{\infty}(M\times M),
\]
the three operators in the composition above having compactly (hence properly) supported kernels. 

{\em Step 2.} We check that we can apply the composition rule for wave front sets.  Note first that since $T$, $T^{-1}$ are (matrices of) pseudo-differential  operators, we obtain using hypothesis (1b) that:
\begin{equation}
\label{wavefr}
_{\rr^{d}}\!\WF(\tilde{c})'= \WF(\tilde{c})'_{\rr^{d}}= \emptyset.
\end{equation}
Let us fix $\alpha, \beta\in \{+,-\}$ and set:
 \[
K_{\alpha}=  \chi (\tilde \rho_{\alpha}\circ E)^{*} \psi_{1},\ K_{\alpha\beta}= \psi_{2}\tilde{c}_{\alpha\beta}\psi_{2}, \ K_{\beta}=  \psi_{1}(\tilde{\rho}_{\beta}\circ E)\chi.
\]  Using (\ref{e4.1}), (\ref{e4.1bis}) we have:
\[
\WF(K_{\alpha})'\subset {\rm Exch}(\Gamma_{\alpha}), \ \WF(K_{\alpha\beta})'\subset \WF(\tilde{c}_{\alpha\beta})', \ \WF(K_{\beta})'\subset \Gamma_{\beta}.
\]
It follows also from (\ref{wavefr}) that:
\[\begin{array}{rl}
&_{M}\!\WF(K_{\alpha})'= \WF(K_{\alpha})'_{\rr^{d}}= _{\rr^{d}}\!\WF(K_{\alpha\beta})'\\[2mm]
&= \WF(K_{\alpha\beta})'_{\rr^{d}}= _{\rr^{d}}\!\WF(K_{\alpha\beta})'= \WF(K_{\beta})'_{M}= \emptyset.
\end{array}
\]
It follows that we can compose $K_{\alpha\beta}$ and $K_{\beta}$ and
\[
\WF(K_{\alpha\beta}\circ K_{\beta})'\subset \WF(\tilde{c}_{\alpha\beta})\circ \Gamma_{\beta}.
\]
We can also compose $K_{\alpha}$ and $K_{\alpha\beta}\circ K_{\beta}$ and 
\[
\WF(K_{\alpha}\circ K_{\alpha\beta}\circ K_{\beta})\subset  {\rm Exch}(\Gamma_{\alpha})\circ \WF(\tilde{c}_{\alpha\beta})\circ \Gamma_{\beta}.
\]
 {\em Step 3.}
 Recalling the definition of $\Gamma_{\alpha}$, $\Gamma_{\beta}$, we obtain from Step 2  that:
\begin{equation}
\label{e4.2}
\begin{array}{rl}
& \WF(C_{\alpha\beta})'\\[2mm]
\subset&\{(x_{1}, \xi_{1}, x_{2}, \xi_{2})\ : \ (x_{1}, \xi_{1})\in \cN_{\alpha}, \ (x_{2}, \xi_{2})\in \cN_{\beta}, \ \exists \ (\rx_{1}, k_{1}, \rx_{2}, k_{2})\in \WF(\tilde{c}_{\alpha\beta})'\\[2mm]
&\hbox{such that }(x_{1}, \xi_{1})\sim (0,\rx_{1}, \alpha\epsilon(0, \rx_{1}, k_{1}),k_{1}), \ (x_{2}, \xi_{2})\sim (0,\rx_{2}, \beta\epsilon(0, \rx_{2}, k_{2}))\}.
\end{array}\end{equation}
Let $S_{\alpha\beta}$ be the set in the r.h.s. of (\ref{e4.2}).  Using (\ref{e4.04}) and the fact that the $S_{\alpha\beta}$ are pairwise disjoint, we obtain that:
\beq\label{astu}
\WF(C)'\subset  \bigcup_{\alpha, \beta \in \{+,-\}}S_{\alpha\beta}.
\eeq
{\em Step 4.}
Recall that we set $\Lambda_{+}= C$, $\Lambda_{-}= C- \i E$.  We first consider the condition 
\[
(\mu{\rm sc}_{+}):\  \WF(\Lambda_{+})'\subset \{(X_{1}, X_{2})\ : \ (X_{i})\in \cN_{+}, \ X_{1}\sim X_{2}\}.
\]
By (\ref{astu}) $(\mu{\rm sc}_{+})$ is satisfied iff:
\[
\begin{array}{rl}
S_{\alpha\beta}= &\emptyset,\hbox{ for }(\alpha, \beta)\neq (+,+),\\[2mm]
S_{++}\subset &\{(x_{1}, \xi_{1}, x_{2}, \xi_{2})\ : \ (x_{i}, \xi_{i})\in \cN_{+}, \ (x_{1}, \xi_{1})\sim (x_{2}, \xi_{2})\}.
\end{array}
\]
This condition is satisfied iff
\beq\label{plus}
\WF(\tilde{c}_{\alpha\beta})'=\emptyset \hbox{ for }(\alpha, \beta)\neq (+, +), \ \WF(\tilde{c}_{++})'\subset \Delta,
\eeq
where $\Delta$ is the diagonal in $\cooo{\rr^{d}}\times \cooo{\rr^{d}}$.
Let us now consider 
\[
(\mu{\rm sc}_{-}):\  \WF(\Lambda_{-})'\subset \{(X_{1}, X_{2})\ : \ (X_{i})\in \cN_{-}, \ X_{1}\sim X_{2}\}.
\]
By (\ref{e0.2}), replacing $C$ by $C-\i E$ amounts to replace  $\tilde{c}$ by $\tilde{c}- \tilde{q}$.  Therefore  $(\mu{\rm sc}_{-})$ is satisfied iff:
\beq\label{minus}
\WF(\tilde{c}_{\alpha\beta}- \delta_{\alpha\beta}\one)'=\emptyset \hbox{ for }(\alpha, \beta)\neq (-, -), \ \WF(\tilde{c}_{--}+\one)'\subset \Delta.
\eeq
 Combining (\ref{plus}) and (\ref{minus}) we obtain that $\Lambda_{\pm}$ satisfy $(\mu{\rm sc})$ iff
 \[
\WF(\tilde{c}_{--})'= \WF(\tilde{c}_{+-})'= \WF(\tilde{c}_{-+})'=  \WF(\one -\tilde{c}_{++})'= \emptyset,
\]
which completes the proof of the theorem.   \qed
\subsection{Construction of Hadamard states}\label{sec4.3}
In this subsection we construct a large class of two-point functions $\lambda$ with pseudo-differential  entries, such that $\Lambda$ is  the two-point function  of a (gauge-invariant) quasi-free Hadamard state.   
Beside the microlocal condition in Thm. \ref{4.2}, $\lambda$ should also satisfy the positivity conditions recalled in Subsect. \ref{sec0.1}, i.e. $\lambda\geq 0$, $\lambda\geq q$, where $q=\i\sigma=\left(\begin{smallmatrix}0 & \one \\ \one & 0 \end{smallmatrix}\right)$.

As before we fix $r\in \cM$.
\begin{proposition}\label{4.4}
Let $\lambda$ be a two-point function with pseudo-differential  entries.  Let $\tilde{\lambda}_{\alpha \beta}$ for $\alpha, \beta\in \{+,-\}$ be defined as in (\ref{e4.05}).
Then $\lambda$ is a Hadamard charge density iff
\[
\begin{array}{rl}
&(\mu{\rm sc}')\ \tilde \lambda_{--}, \ \tilde \lambda_{+-}, \tilde \lambda_{-+}\in \Psi^{-\infty}(\rr^{d}),\\[2mm]
&(1)\ \tilde \lambda_{++}\geq \one, \hbox{ on }\cH(\rr^{d}), \ \tilde \lambda_{--}\geq 0\hbox{ on }\cH(\rr^{d}),\\[2mm]
&(2)\ |(u| \tilde \lambda_{+-}v)|\leq (u|\tilde  \lambda_{++}u)^{\12}(v|\tilde  \lambda_{--}v)^{\12}, \ u, v\in \cH(\rr^{d}),\\[2mm]
&(3)\ |(u| \tilde\lambda_{+-}v)|\leq (u| (\tilde\lambda_{++} -\one)u)^{\12}(v|(\tilde\lambda_{--}+\one)v)^{\12}, \ u, v\in \cH(\rr^{d}).
\end{array}
\]
\end{proposition}
\proof 
Since the entries of $\lambda$, $\tilde \lambda$ are pseudo-differential  operators, condition (1a) of Thm. \ref{4.2} is satisfied and condition (1b) as well by Remark \ref{remaro}. Moreover, condition $(\mu{\rm sc}')$ is equivalent to (2) of Thm. \ref{4.2}, hence $(\mu{\rm sc}')$ is equivalent to the microlocal spectrum condition.	

From Sect. \ref{sec0} we know that  $\lambda$ is the two-point function of a gauge-invariant quasi-free state iff 
\beq\label{e4.7}
 \lambda\geq 0, \ \lambda\geq q \ \ \hbox{on }\cD\shit
\Leftrightarrow  \lambda\geq 0, \ \lambda\geq q \ \  \hbox{on }\cH\shit,
\eeq
using that  the entries of $\lambda$ are  pseudo-differential operators.

We recall that  $\tilde{\lambda}= (T^{-1})^{*}\circ \lambda \circ T^{-1}$ and  $\tilde{q}= (T^{-1})^{*}\circ q \circ T^{-1}$.   By Lemma \ref{4.1} we have 
\[
\tilde{q}= \mat{\one}{0}{0}{ -\one}: \ \cH\shit \to \cH'\shit.
\]   

Since $T$ maps $\cH\shit$ into itself bijectively, (\ref{e4.7}) is equivalent to:
\begin{equation}
\label{e4.3}
\tilde{\lambda}\geq 0, \  \tilde{\lambda}\geq \tilde{q} \ \ \hbox{on }\cH\shit.
\end{equation} 
Clearly if $a, b, c$ are linear operators with domain $\cH(\rr^{d})$ one has:
\[
\begin{array}{rl}
&2{\rm Re}(u| bv)+ (u|au)+ (v| cv)\geq 0, \ u, v\in \cH(\rr^{d})\\[2mm]
\Leftrightarrow& |(u| bv)|\leq (u| au)^{\12}(v|cv)^{\12}, \ u, v\in \cH(\rr^{d})\hbox{ and }a, c\geq 0\hbox{ on }\cH(\rr^{d}).
\end{array}
\]
Applying this observation and noting that $r+r^{*}\geq 0$, we obtain that condition (\ref{e4.3}) is equivalent to conditions (1), (2), (3). \qed

We now proceed to construct a large class of pseudo-differential  operators $\tilde\lambda_{\alpha\beta}$ satisfying the conditions in Prop. \ref{4.4}.

\bet\label{4.3}
Let us fix pseudo-differential  operators:
\[
a_{-\infty}, b_{-\infty}\in \Psi^{-\infty}(\rr^{d}), \ a_{0}\in \Psi^{0}(\rr^{d}) \hbox{ with }\|a_{0}\|\leq 1,
\]
and set:
\[
\begin{array}{rl}
\tilde{\lambda}_{++}=&\one+ b_{-\infty}^{*}b_{-\infty},\\[2mm]
\tilde \lambda_{--}= &a_{-\infty}^{*}a_{-\infty},\\[2mm]
\tilde \lambda_{+-}=&\tilde \lambda_{-+}^{*}= b_{-\infty}^{*}a_{0}a_{-\infty}
.\end{array}
\]
Then the two-point function $\lambda$ given by (\ref{e4.05}) is the two-point function of a Hadamard state. 
\eet

\proof  We  check the conditions in Prop. \ref{4.4}.
Conditions $(\mu{\rm sc})$ and (1) are clearly satisfied.   From the form of $\tilde{\lambda}_{+-}$ we have
\[
|(u| \tilde{\lambda}_{+-}v)|\leq  (u| b_{-\infty}^{*}b_{-\infty}u)^{\12}(v| a_{-\infty}^{*}a_{-\infty}v)^{\12}, \ u,v\in \cH(\rr^{d}),
\]
which implies (2) and (3), using the form of $\tilde{\lambda}_{++}$ and $\tilde{\lambda}_{--}$.  \qed

\subsection{Symplectic transformations}\label{sec4.4}
Recall from Sect. \ref{sec0} that if $(\cY, \sigma)$ is a complex symplectic space and $q= \i \sigma$, then the set of two-point functions of gauge-invariant quasi-free states is invariant under conjugation by elements of $U(\cY, q)$. The same is true for the set of two-point functions of pure quasi-free states.

In this subsection we describe a class of symplectic transformations $u\in U(\cD(\rr^{d})\otimes \cc^{2},q)$ which preserve the  microlocal spectrum condition $(\mu{\rm sc})$. We start with a general result.

\begin{proposition}\label{p4.4}
 Let
 $u$ such that $u, u^{*}: \cH(\rr^{d})\otimes \cc^{2}\to \cH(\rr^{d})\otimes \cc^{2}$.
 Set
 \[
\tilde{u}:= TuT^{-1}=\mat{\tilde{u}_{++}}{\tilde{u}_{+-}}{\tilde{u}_{-+}}{\tilde{u}_{--}},
\]
and assume that 
\[
\WF(\tilde{u}_{++}^{*} \tilde{\lambda}_{++}\tilde{u}_{++})'\subset \Delta, \ \tilde{u}_{+-}: \cH'(\rr^{d})\to \cH(\rr^{d}).
\]
Then if $\lambda$ is a two-point function with pseudo-differential  entries satisfying $(\mu{\rm sc})$,  the two-point function 
$u^{*}\lambda u$ satisfies also $(\mu{\rm sc})$. 
\end{proposition}
\proof
We set  $ c:= u^{*}\lambda u$ and as in Subsect. \ref{sec4.2}:
\[
\tilde{\lambda}:= (T^{-1})^{*} \lambda T^{-1}, \ \tilde{c}:= (T^{-1})^{*} c T^{-1}= \tilde{u}^{*} \tilde{\lambda}\tilde{u}.
\]
Since $\lambda$ has pseudo-differential  entries and satisfies $(\mu{\rm sc})$ we have:
\beq\label{e.buda7}
\tilde{\lambda}_{\alpha\beta}\in \Psi^{\infty}(\rr^{d}), \ \tilde{\lambda}_{\alpha\beta}\in \Psi^{-\infty}(\rr^{d}) \ \hbox{ for }(\alpha, \beta)\neq (+,+).
\eeq

We will check that $c$ satisfies the hypotheses of Thm. \ref{4.2}. Since $u, u^{*}, \lambda$ preserve $\cH(\rr^{d})\otimes \cc^{2}$ condition (1a) is satisfied. By Remark \ref{remaro}, it remains to check condition (2). We compute $\tilde{c}$ and 
obtain using (\ref{e.buda7}) that
\[
\tilde{c}=\mat{\tilde{u}_{++}^{*}\tilde{\lambda}_{++}\tilde{u}_{++}}{\tilde{u}^{*}_{++}\tilde{\lambda}_{++}\tilde{u}_{+-}}{\tilde{u}^{*}_{+-}\tilde{\lambda}_{++}\tilde{u}_{++}}{\tilde{u}^{*}_{+-}\tilde{\lambda}_{++}\tilde{u}_{+-}}+ s,
\]
 where $s: \cH'(\rr^{d})\otimes \cc^{2}\to \cH(\rr^{d})\otimes \cc^{2}$ is a smoothing operator.  Since $\tilde{u}_{+-}, \tilde{u}_{+-}:\cH'(\rr^{d})\to \cH(\rr^{d})$ we have 
 \[
\tilde{c}=\mat{\tilde{u}_{++}^{*}\tilde{\lambda}_{++}\tilde{u}_{++}}{0}{0}{0}+ s_{1},
\]
for $s_{1}$ as $s$. Therefore condition (2) is satisfied. \qed
\medskip

\begin{definition}
 We denote by $U_{-\infty}(\cH(\rr^{d})\otimes \cc^{2}, q)$  the subgroup of $U(\cH(\rr^{d})\otimes \cc^{2}, q)$ defined by:
\[
U_{-\infty}(\cH(\rr^{d})\otimes \cc^{2}, q):=\{u\in U(\cH(\rr^{d})\otimes \cc^{2}, q)\ : \ u-\one\in\Psi^{-\infty}(\rr^{d})\otimes M_{2}(\cc)\}.
\]
\end{definition}
\begin{corollary}\label{p4.5}
The conjugations by elements of $U_{-\infty}(\cH(\rr^{d})\otimes \cc^{2}, q)$ preserve the set of (pure) quasi-free Hadamard states. \end{corollary}

\begin{remark}
It is easy to see that 
 if $\mat{a}{b}{c}{d}\in U(\cH(\rr^{d})\otimes \cc^{2}, q)$ and  $a$ is invertible, then
\beq
\mat{a}{b}{c}{d}=\mat{\one}{0}{e}{\one}\mat{g^*}{0}{0}{g^{-1}}\mat{\one}{f}{0}{\one}
\eeq 
for some $g$ invertible and $e^*=-e$, $f^*=-f$.  Moreover  the matrices 
\[
(1) \ \mat{\one}{0}{e}{\one}, \ \ (2) \  \mat{\one}{f}{0}{\one} \mbox{ or } \ (3) \  \mat{g^*}{0}{0}{g^{-1}},
\]
where $e, \ f\in \Psi^{-\infty}(\rr^{d})$ with $e^*=-e$, $f^*=-f$,  and $g-\one\in \Psi^{-\infty}(\rr^{d})$ with $g$, $g^{*}$  invertible, belong to $U_{-\infty}(\cH(\rr^{d})\otimes \cc^{2}, q)$.
\end{remark}
\

\subsection{Pure Hadamard states}\label{sec4.5}
We now characterize pure Hadamard states with pseudo-differential  entries and discuss some examples. 

%
%

\begin{theoreme}\label{thm:canonicform}Let $\lambda\in\Psi^{\infty}(\rr^d)\otimes M_2(\cc)$. Then $\lambda$ is the two-point function of a pure Hadamard state iff
\beq\label{eq:canonicform}
\begin{array}{rl}
\tilde{\lambda}_{++}=&\one+a_{-\infty}a_{-\infty}^*,\\[2mm]
\tilde \lambda_{--}= &a_{-\infty}^{*}a_{-\infty},\\[2mm]
\tilde \lambda_{+-}=&\tilde \lambda_{-+}^{*}= a_{-\infty}(\one+a_{-\infty}^*a_{-\infty})^{\half}
\end{array}
\eeq
for some $a_{-\infty}\in\Psi^{-\infty}(\rr^d)$.
\end{theoreme}
\proof 
Set $\tilde\eta= 2\tilde \lambda-\tilde q$. From Prop. \ref{istu} we see that $\lambda$ is the two-point function of a pure state iff
\begin{equation}
\label{turu}
i) \ \tilde\eta\geq 0, \ ii) \ \tilde \eta \tilde q^{-1} \tilde \eta=\tilde q.
\end{equation}
Writing $\tilde\eta$ as $\mat{a}{b}{b^{*}}{c}$ we obtain that (\ref{turu}) is equivalent to:
\begin{equation}
\label{toro}
\begin{array}{rl}
i')& a\geq0, \ c\geq 0, \ | (u| bv)|\leq  (u| au)^{\12}(v| cv)^{\12}, \ u, v\in \cH(\rr^{d}),\\[2mm]
ii')& a^{2}= \one + bb^{*}, \ c^{2}= \one + b^{*}b, \ ab-bc=0.
\end{array}
\end{equation}
Note that if $b$ is a bounded operator on $L^{2}(\rr^{d})$ then:
\begin{equation}
\label{tara}
bf(b^{*}b)= f(bb^{*})b, \hbox{ for any Borel function }f.
\end{equation}
In fact (\ref{tara}) is immediate for $f(\lambda)= (\lambda-z)^{-1}$, $z\in \cc\backslash \rr$ and extends to all Borel functions by a standard argument.  

Since $a,c\geq 0$ by {\it i')}, the first two equations of {\it ii')} yield 
\[
a= (\one + bb^{*})^{\12}, \ c= (\one + b^{*}b)^{\12}.
\] The third equation of {\it ii')} then holds using (\ref{tara}). The second condition in {\it i')} is equivalent to $\| (\one + bb^{*})^{\12}b(\one + b^{*}b)^{\12}\|\leq 1$, which holds using again (\ref{tara}).

 Going back to $\tilde\lambda$ we obtain
\begin{equation}
\label{tiri}
\tilde\lambda=\12\mat{(\one + bb^{*})^{\12}+ \one}{b}{b^{*}}{(\one + b^{*}b)^{\12}-\one}.
\end{equation}
Let now 
\[
a:= \frac{b}{\sqrt{2}}((\one + b^{*}b)^{\12}+ \one)^{\12}.
\]
Using (\ref{tara}) we obtain by an easy computation that
\[
\one + a^{*}a= \12((\one + b^{*}b)^{\12}+ \one), \ \one + aa^{*}=  \12((\one + bb^{*})^{\12}+ \one), \ b= 2a(\one + a^{*}a)^{\12}.
\]
Hence $\tilde\lambda$ in (\ref{tiri}) can be rewritten as:
\begin{equation}
\label{tere}
\tilde\lambda= \mat{\one + aa^{*}}{a(\one + a^{*}a)^{\12}}{(\one + a^{*}a)^{\12}a^{*}}{a^{*}a}.
\end{equation}
By Prop. \ref{4.4} $\lambda$ satisfies $(\mu{\rm sc})$ iff $a^{*}a$, $a(\one + a^{*}a)^{\12}\in\Psi^{-\infty}$, which is equivalent to $a\in \Psi^{-\infty}$. \qed

\begin{proposition}\label{prop:transinfty}
Let $\lambda_i\in\Psi^{\infty}(\rr^d)\otimes M_2(\cc)$, $i=1,2$, be two-point functions of pure Hadamard states (for the model Klein-Gordon equation). Then there exists $u\in U_{-\infty}(\cH(\rr^{d})\otimes \cc^{2}, q)$ s.t.
\[
\lambda_2=u^* \lambda_1 u.
\]
\end{proposition}
\proof Without loss of generality we can assume that $\tilde\lambda_{1}= \mat{\one}{0}{0}{0}$ and $\tilde{\lambda}_{2}$ is given by (\ref{tere}) for $a\in \Psi^{-\infty}$.  Then
\[
\tilde{u}=\mat{(\one + aa^{*})^{\12}}{a}{a^{*}}{(\one + a^{*}a)^{\12}}
\]
belongs to $U_{\infty}(\cH(\rr^{d})\otimes \cc^{2},\tilde q)$ and satisfies $\tilde{u}^{*}\tilde{\lambda}_{1}\tilde{u}= \tilde{\lambda}_{1}$. \qed

\subsubsection{Canonical Hadamard state}\label{ssec:canonic}

Once having fixed $r\in\cM$, let us consider the two-point function 
\[
\lambda(r):=\mat{(r+r^{*})^{-1}}{(r+r^{*})^{-1}r}{r^{*}(r+r^{*})^{-1}}{r^{*}(r+r^{*})^{-1}r}.
\]
An easy computation shows that
\[
\tilde\lambda(r)= (T(r)^{-1})^* \lambda(r) T(r)^{-1} = \mat{\one}{0}{0}{0}.
\]
This is a particular case of Theorem \ref{thm:canonicform} with $a_{-\infty}=0$ and it follows that $\lambda(r)$ is the two-point function of a pure Hadamard state. One can show that it is distinguished among all two-point functions $\lambda:\cH\shit\to\cH\shit$ of pure quasi-free states by the property
\[
{\rm Ran}\, P_{\pm} \subset C_{\pm}(r), 
\]
where $P_{\pm}$ is defined on $H^{1}(\rr^d)\oplus L^{2}(\rr^d) $ by
\[
P_{\pm}:=\shalf\one\pm q\eta, \quad \eta=\lambda-\half q.
\]

We now study the dependence of  $\lambda(r)$ on $r\in\cM$.

\begin{proposition}\label{defu}
Let:
 \[
\cG:=\{(g, f)\ : \ g-\one, \ f\in \Psi^{-\infty}(\rr^{d}), \  g, g^{*}:L^{2}(\rr^{d})\to L^{2}(\rr^{d})\hbox{ invertible, }f=-f^{*}\}.
\]
We equip $\cG$ with the group structure given by:
\[
\begin{array}{rl}
&{\rm Id}:= (\one, 0),\\[2mm]
&G_{2}G_{1}:=(g_{2}g_{1}, (g_{2}^{*})^{-1}f_{1}g_{2}^{-1}+ f_{2}), \hbox{ for } G_{i}= (g_{i}, f_{i}).
\end{array}
\]
Then the following holds:
\ben
\item the map 
\[
\cG\ni G=(g,f)\mapsto u_{G}:=  \mat{g^*}{0}{0}{g^{-1}}\mat{\one}{f}{0}{\one}\in U_{-\infty}(\cH(\rr^{d})\otimes \cc^{2}, q)
\]
is a group homomorphism. 
\item 
\
 $\cG$ acts transitively on $\cM$ by
 \[
\alpha_{G}(r):= (g^{*})^{-1} r g^{-1}+f, \ r\in \cM, \ G= (g,f)\in \cG.
\]
\een
\end{proposition}
\proof  Statement (1) is an easy computation. Let us prove (2).

We  first check that $\alpha_{G}$ preserves $\cM$. Let $r\in \cM$ and $\tilde{r}= \alpha_{G}(r)$ for $G\in \cG$. Clearly  $\tilde{r}-r\in \Psi^{-\infty}$ so $\tilde{r}- (b^{*})^{(-1)}= r-(b^{*})^{(-1)}+ \Psi^{-\infty}\in \Psi^{-\infty}$.  It remains to check that
\begin{equation}
\label{etoto.10}
\tilde{r}+ \tilde{r}^{*}\sim  \epsilon^{-1}.
\end{equation}
This is obvious if $G= (\one, f)$, since then $\tilde{r}= r+f$ and $f^{*}=-f$. 

Assume now that $G= (g, 0)$, so that $\tilde{r}+ \tilde{r}^{*}= (g^{*})^{-1}(r+ r^{*})g^{-1}$, and $g-\one \in \Psi^{-\infty}$, $g, g^{*}: L^{2}\to L^{2}$ invertible.
It follows that
\beq\label{etoto.11}
(\tilde{r}+ \tilde{r}^{*})^{-1}= g (r+ r^{*})^{-1}g^{*}= \12 \epsilon + \Psi^{0}.
\eeq
Since $r\in \cM$ we have $r+ r^{*} \sim  \epsilon^{-1}$, hence $(r+ r^{*})^{-1} \sim  \epsilon$, by the Kato Heinz inequality.
In particular we have $(r+ r^{*})^{-1}\geq c_{0}>0$. Using then (\ref{etoto.11}) we obtain that:
\[
\begin{array}{rl}
(\tilde{r}+ \tilde{r}^{*})^{-1}\geq&  c_{3}>0,\\[2mm]
 (\tilde{r}+ \tilde{r}^{*})^{-1}\geq& \12 \epsilon- c_{4}.
\end{array}
\]
This implies that $(\tilde{r}+ \tilde{r}^{*})^{-1}\geq c \epsilon$ for some $c>0$. On the other hand (\ref{etoto.11}) directly implies that $(\tilde{r}+ \tilde{r}^{*})^{-1}\leq c \epsilon$ for some $c>0$. Therefore we have $(\tilde{r}+ \tilde{r}^{*})^{-1}\sim  \epsilon$, which implies (\ref{etoto.10}) by applying Kato-Heinz theorem once again.  This completes the proof that $\alpha_{G}$ preserves $\cM$. 

It remains to prove that the action is transitive. 

 let $r_{i}\in \cM$, $i=1,2$  As we saw above  $(r_{i}+ r_{i}^{*})^{-1}\in \Psi^{1}$  and $(r_{i}+ r_{i}^{*})^{-1}\sim  \epsilon$.  By Prop. \ref{1.1} we obtain that $(r_{i}+r_{i}^{*})^{-\12}\in\Psi^{\12}$ and by Kato-Heinz theorem we have $(r_{i}+r_{i}^{*})^{-\12}\sim  \epsilon^{\12}$. In particular $(r_{i}+r_{i}^{*})^{-\12}$ is bijective from $H^{\12}(\rr^{d})$ to $L^{2}(\rr^{d})$. It follows that
\beq\label{etiti.2}
(r_{2}+r_{2}^{*})^{-\12}= g(r_{1}+r_{1}^{*})^{-\12}= (r_{1}+r_{1}^{*})^{-\12}g^{*},
\eeq
where $g$, $g^{*}$ are invertible on $L^{2}(\rr^{d})$. Using also that $r_{1}-r_{2}\in \Psi^{-\infty}$, we obtain that $g-\one\in \Psi^{-\infty}$.
From (\ref{etiti.2}) we get:
\begin{equation}
\label{etiti.3}
r_{2}+r_{2}^{*}= (g^{*})^{-1}(r_{1}+r_{1}^{*})g^{-1}.
\end{equation}
We set now
\beq\label{etiti.4}
 r_{2}- r_{2}^{*}=: (g^{*})^{-1}(r_{1}- r_{1}^{*})g^{-1}+2f. 
\eeq
Clearly $f^{*}= -f$, and since $g-\one$ and $r_{1}-r_{2}$ belong to $\Psi^{-\infty}$, we see that $f\in \Psi^{-\infty}$. From (\ref{etiti.3}), (\ref{etiti.4}) we obtain that $r_{2}= (g^{*})^{-1}r_{1}g^{-1}+f= \alpha_{G}(r_{1})$,  for $G= (g,f)$. This completes the proof of the proposition. \qed

\medskip

The following theorem explains the dependence of the pure quasi-free state with two-point function $\lambda(r)$ on the choice of $r\in \cM$.
\begin{theoreme}\label{p4.7}
We have 
\[
\lambda(\alpha_{G}(r))= u_{G}^{*}\lambda(r)u_{G}, \ \forall \ r\in \cM, \ G\in \cG.
\]

\end{theoreme}
\proof  writing $\lambda(r)$ as:
\[
 \lambda(r)= \mat{\one}{0}{r^{*}}{\one}\mat{(r+r^{*})^{-1}}{0}{0}{0}\mat{\one}{r}{0}{\one},
\]
we easily obtain that if $u=\mat{\one}{f}{0}{\one}$, then
\[
u^{*}\lambda(r)u= \lambda(r+f),
\]
and if $u= \mat{g^{*}}{0}{0}{g^{-1}}$, then
\[
u^{*}\lambda(r)u= \lambda((g^{*})^{-1}rg^{-1}).
\]
This completes the proof of the theorem. \qed

\subsection{The static case}\label{for-record}
Let us  illustrate our results in  the {\em static case}, when $a(t, \rx, D_{\rx})$ is independent on $t$. We assume for simplicity that $a(\rx, D_{\rx})\geq m^{2}>0$, in order to avoid infrared problems.   We can work in an abstract setting and denote by $a>0$ a selfadjoint operator on a (complex) Hilbert space $\ch$.   We set $\epsilon:= a^{\12}$. 

The solution of the Cauchy problem:
\beq\label{e5.1}
\left\{
\begin{array}{rl}
&\p_{t}^{2}\phi(t)+a\phi(t)=0,\\[2mm]
&\phi(0)=f_{0},\\[2mm]
&\i^{-1}\p_{t} \phi(0)=f_{1},
\end{array}
\right.
\eeq

is:
 \[
\phi(t)= \shalf\e^{\i t \epsilon}(f_{0}+ \epsilon^{-1}f_{1})+ \shalf\e^{-\i t \epsilon}( f_{0}- \epsilon^{-1}f_{1})=:U(t,0)f.
\]
Therefore, when $\ch=L^2(\rr^d)$ and $\epsilon\in\Psi^{1}(\rr^d)$ we can choose 
\[
b(t)= \epsilon,  \ u_{\pm}(t,s)= \e^{\pm \i (t-s)\epsilon}, \ d_{\pm}(s)= \12\one, \ r_{\pm}(s)= \epsilon^{-1}.
\]
\begin{remark}Using the reduction to the model case described in Subsect. \ref{ss:modelcase}, one obtains $a(t, \rx, D_{\rx})$ independent on $t$ if the metric is static and the electric field vanishes, i.e. $\partial_i V+\partial_t A_i\equiv 0$, $i=1,\dots, d$.
\end{remark}

For sake of completeness we list below the essential examples of Hadamard states in the static case.  
\begin{itemize}
\item The two-point function of the vacuum state is:
\[
(f|\lambda_{\rm vac}f)= \12(f_{0}+ \epsilon^{-1}f_{1}|\epsilon(f_{0}+\epsilon^{-1}f_{1}))_{\ch}.
\]
The matrix elements of $\tilde{\lambda}_{\rm vac}$ are:
\[
\tilde\lambda_{++}= \one, \ \ \tilde \lambda_{--}= \tilde \lambda_{+-}= \tilde \lambda_{-+}=0.
\]
It follows that $\lambda_{\rm vac}$ equals to $\lambda(\epsilon^{-1})$ with the notation in Subsect. \ref{ssec:canonic}. Setting 
\[
\phi_+(t):=U(t,0)P_+ f=U(t,0)(q\lambda ) f 
\]
we have
\[
\phi_+(t)=\shalf\e^{\i t \epsilon}( f_{0}+  \epsilon^{-1}f_{1}).
\]

\item Let us consider a special case of Theorem \ref{4.3}, namely let $\lambda$ be such that the entries of $\tilde\lambda$ are given by
\[ 
\tilde{\lambda}_{++}=\one+ b, \ \ \tilde \lambda_{+-}=\tilde \lambda_{-+}^{*}= 0, \ \ \tilde \lambda_{--}= a,
\]
where $a, b\in\Psi^{-\infty}(\rr^d)$ are both assumed to be positive. The corresponding state is not pure unless $a=b=0$. More explicitly, $\lambda$ is given by
\[
\lambda=\half\mat{\epsilon+\epsilon^{\half}(a+b)\epsilon^{\half}}{\one+\epsilon^{\half}(b-a)\epsilon^{-\half}}{\one+\epsilon^{-\half}(b-a)\epsilon^{\half}}{\epsilon^{-1}+\epsilon^{-\half}(a+b)\epsilon^{-\half}}
\]
Defining $\phi_+(t)$ as before we get
\begin{eqnarray*}
\phi_+(t)= & \shalf\e^{\i t \epsilon}\left((\one+\epsilon^{-\half}b\epsilon^{\half})f_{0}+ \epsilon^{-1}(\one+\epsilon^{-\half}b\epsilon^{\half})f_{1}\right)
 \\ & + \shalf\e^{-\i t \epsilon}\left(\epsilon^{-\half}a\epsilon^{\half} f_{0}- \epsilon^{-\half}a\epsilon^{-\half} f_{1}\right).
\end{eqnarray*}
One can show that the thermal state at inverse temperature $\beta$ is obtained by taking
\[
a=b=\frac{\e^{- \beta \epsilon}}{\one - \e^{- \beta \epsilon}}.
\]
\end{itemize}

\section{Hadamard states on general space-times}\label{sec7}\init

\subsection{Space-times with compact Cauchy surfaces}\label{ssec:junker}
The results  in Sects. \ref{sec1}, \ref{sec2}, \ref{sec3}  and \ref{sec4} extend verbatim to the case where $\rr^{d}$ is replaced by a compact manifold $S$. It suffices to replace everywhere  $\cE(\rr^{d})$,  $\cH(\rr^{d})$ and $\cD(\rr^{d})$ by $\cD(S)$ and similarly for their dual spaces.  The Weyl pseudo-differential  calculus has to be replaced by the  standard calculus on compact manifolds.
This case is related to the results in \cite{junker,JS02}.

\begin{remark}In \cite{junker,JS02}, a different convention is employed for the symplectic form acting on Cauchy data. This amounts to considering $(\phi(s),\partial_t\phi(s))$ as Cauchy data instead of $(\phi(s),\i^{-1}\partial_t\phi(s))$. A two-point function $\lambda_{\rm Ju}$ in the convention used in \cite{junker,JS02} corresponds in our notation to the two point function $\lambda=v\lambda_{\rm Ju}v^*$, where $v$ is  diagonal  with entries $v_{++}=\one$ and $v_{--}=\i \one$.
\end{remark}

In \cite[Thm. 5.10]{JS02} it is shown how to construct families of operators $J(t)\in C^{\infty}(\rr,\Psi^1(S))$, $R(t)\in C^{\infty}(\rr,\Psi^0(S))$, such that
\[
\lambda=\half\mat{RJ^{-1}R+J}{\one-\i R J^{-1}}{\one+\i J^{-1}R}{J^{-1}}
\]
is the two-point function of a pure Hadamard state on $\rr\times S$. In our approach, this corresponds to setting
\[
R(t)=\frac{\i}{2} \left(b(t) - b^*(t)\right), \quad J(t)=\frac{1}{2} \left(b(t) + b^*(t)\right),
\]
where $b(t)$ is as in Theorem \ref{3.1}. Using $r(t)=b^*(t)^{(-1)}$ mod $C^{\infty}(\rr,\Psi^{-\infty})$, it is not difficult to check the microlocal spectrum condition by means of Theorem \ref{4.2}. It is worth pointing out that one of the advantages of basing the construction on $r(t)$ (as we do) rather than on $b(t)$ is that the former is more closely related to the operator $\epsilon(t)$, cf. Theorem \ref{3.1}.

\begin{remark}
 Since $S$ is compact we know that $\Op(a): H^{m}(S)\to H^{p}(S)$ is Hilbert-Schmidt for any $m,p\in \rr$  and $a\in \Psi^{-\infty}(\rr^{d})$.
 By Shale's theorem it follows that the CCR representations obtained from two pure Hadamard states as in Thm. \ref{thm:canonicform} are unitarily equivalent, since two such states are obtained from one another by a symplectic transformation in $U_{\infty}(\cD(S)\otimes \cc^{2},q)$.
\end{remark}
\subsection{General space-times}\label{ssec:general}

In this subsection we give a new construction of quasi-free Hadamard states on an arbitrary globally hyperbolic space-time $M$ and compare it with the classical construction of Fulling, Narcowich and Wald \cite{FNW}. For the reader's convenience let us first outline this construction.
 \subsubsection{The FNW construction}
Let $(M,g)$ a globally hyperbolic space-time, $S$ a Cauchy hypersurface.  We may assume that 
\[
M= \rr_{t}\times S, \quad g= -c(x)dt^{2}+ h_{jk}(x)d\rx^{j}d\rx^{k}
\]
and we set $S_{t}:= \{t\}\times S$. We fix a real function $r\in \cinf(M)$ and consider $P= - \nabla^{a}\nabla_{a}+ r(x)$ --- the associated Klein-Gordon operator. The construction of Hadamard states for $P$ is performed using a `deformation argument'. Namely, one chooses an  ultra-static metric 
\[
g^{0}= - dt^{2}+ h_{jk}(\rx)d\rx^{j}d\rx^{k}, \quad r^{0}(x)= m^{2}>0,
\]
and interpolating  metric $g'= - c'(x)dt^{2}+ h_{jk}'(x)d\rx^{j}d\rx^{k}$, and  real function $r'\in \cinf(M)$ such that  $(g',r')= (g,r)$ near $[-T/2, T/2]\times S$, $(g',r')= (g^{0}, m^{2})$ near $\rr\backslash [-T, T]\times S$.
We denote by $P^{0}, P'$ the associated Klein-Gordon operators, and by $E, E^{0}, E'$ the respective Pauli-Jordan commutator functions.

It is easy to show  that $E'(x, y)= E(x,y)$ (resp. $E'(x,y)=E^{0}(x, y)$) for $x, y$ in a neighborhood of $\{0\}\times S$ (resp. $\{-T\}\times S$).

Let now $\omega_{0}$ be a quasi-free Hadamard state for $P^{0}$. Such Hadamard state always exists on an ultra-static space-time, for instance one can take the associated ground state. Parametrizing elements of ${\rm Sol}_{\rm sc}(P^{0})$ by Cauchy data on $S_{-T}$, we can consider its two point function  $\lambda_{0}$, acting on $\coinf(S_{-T})\otimes \cc^{2}$.  Then  $\Lambda'_{0}:= (\rho_{-T}\circ E')^{*}\circ \lambda_{0}\circ (\rho_{T}\circ E')$ satisfies the Hadamard condition for $P'$ in a neighborhood of $S_{-T}$, since the kernels of $E'$ and $E^{0}$ coincide on a neighborhood of $S_{-T}$ and $\omega^{0}$ is Hadamard for $P^{0}$.  By a well-known argument $\Lambda^{'}_{0}$ satisfies the Hadamard condition for $P'$ globally, and defines a Hadamard state $\omega_{0}'$ for $P'$.

We now associate to $\omega_{0}'$ its two-point function $\lambda_{0}'$ acting on $\coinf(S_{0})\otimes\cc^{2}$, expressed in terms of Cauchy data on $S_{0}$. Then by the same argument $\Lambda:=(\rho_{0}\circ E)^{*}\circ \lambda_{0}'\circ (\rho_{0}\circ E)$  satisfies the Hadamard condition for $P$ locally near  $S_{0}$  hence globally. In this way we obtain a quasi-free Hadamard state $ \omega$ for $P$.

Denoting  by $U'(-T, 0): \cD(S_{0})\otimes\cc^{2}\to \cD(S_{-T})\otimes \cc^{2}$ the propagator for $P'$, mapping the Cauchy data on $S_{0}$ to the Cauchy data on $S_{-T}$ we have:
\beq\label{ascoli}
\lambda_{0}'= U'(-T, 0)^{*}\circ \lambda_{0}\circ U'(-T, 0).
\eeq
Since $U'(-T,0)$ is symplectic we see  by Prop. \ref{istu} that $\omega'_{0}$ is pure iff $\omega_{0}$ is pure, and the same argument gives pureness of $\omega$. 

\subsubsection{An alternative construction}
In our approach, we reduce the general problem to the special case of space-times considered by us in Sections \ref{sec2}-\ref{sec3} (or simply to the case of a compact Cauchy surface). Namely, using a suitable partition of unity, we glue together Hadamard states on smaller regions of the space-time. \medskip
 
The steps of the construction are the following:

\medskip
 
We fix a Cauchy surface $S$, so that we can assume that $M=\rr\times S$ and the metric $g$ is of the form (\ref{metric}).

We choose  an open set $\Omega$ in $M$ and  for $n\in \nn$,  open, pre-compact sets  $U_{n}$, $\tilde{U}_{n}$ in $S$,  constants
 $0<\delta_{n}$    such that:
\begin{equation}
\label{Had.e1}
\begin{array}{rl}
(i)&U_{n}\Subset \tilde{U}_{n}, \ \bigcup_{n}U_{n}=S,\\[2mm]
(ii)&\tilde{U}_{n}\hbox{ are coordinate charts for }S,\\[2mm]
(iii)&y\in \Omega, \ J(y)\cap U_{n}\neq \emptyset \ \Rightarrow \ y\in ]-\delta_{n}, \delta_{n}[\times \tilde{U}_{n}=: \tilde{\Omega}_{n},\\[2mm]
(iv)&\Omega\hbox{ is a neighborhood of }S\hbox{ in }M.
\end{array}
\end{equation}
In {\it (iii)} $J(y)$ denotes the causal shadow of $y\in M$.
This is clearly possible.

We fix a partition of unity $1= \sum_{n}\chi_{n}^{2}$ of $S$, with  $\chi_{n}\in \coinf(U_{n})$ for $n\in \nn$. Denoting still by $\chi_{n}$ the map $\chi_{n}\otimes \one$ on $\coinf(S)\otimes \cc^{2}$, we note that
\begin{equation}
\label{Had.e2}
q= \sum_{n}\chi_{n}^{*}q\chi_{n}.
\end{equation}

Fix for each $n\in \nn$ a coordinate map $\varphi_{n}: \tilde{U}_{n}:\to \tilde{V}_{n}$, 
where $V_{n}$ is a neighborhood of $0$ in $\rr^{d}$.  The symplectic form $\sigma$ 
on $\coinf(\tilde{U}_{n})\otimes \cc^{2}$ transported to $\coinf(\tilde{V}_{n})\otimes\cc^{2}$ will be given by  (\ref{e2.0}).

We also transport  with $\varphi_{n}$ the Klein-Gordon operator $P$ on $\tilde{\Omega}_{n}$ to an operator on $]-\delta_{n}, \delta_{n}[\times V_{n}\subset \rr\times \rr^{d}$.  We can extend this operator to $\rr\times \rr^{d}$ so that it satisfies the conditions in Sect. \ref{sec2}. Let us denote by $P_{n}$ the Klein-Gordon operator on $\rr\times \rr^{d}$ obtained in this way. We choose for each $n\in \nn$ a two-point function $c_{n}$ (acting on the  space of Cauchy data) of a quasi-free state, which is Hadamard for $P_{n}$.  We will have in particular
\beq\label{Had.e3}
c_{n}\geq 0, \ c_{n}\geq q.
\eeq
We restrict $c_{n}$ to $\coinf(\tilde{V}_{n})\otimes \cc^{2}$ and transport it back to $\coinf(\tilde{U}_{n})\otimes \cc^{2}$ by $\varphi_{n}^{-1}$, denoting it by $\lambda_{n}$. Finally we set:
\beq\label{ascola}
\lambda:= \sum_{n\in \nn}\chi_{n}^{*}\lambda_{n}\chi_{n},
\eeq
which is well defined as a two-point function on $\coinf(S)\otimes \cc^{2}$, since the sum is locally finite. 

Since $q= \sum \chi_{n}^{*}q\chi_{n}$, and $c_{n}$ was the two-point function of a gauge-invariant quasi-free state, we deduce from (\ref{Had.e3}) that  $\lambda\geq 0$, $\lambda\geq q$, i.e. $\lambda$ is the two-point function of a gauge-invariant quasi-free state. 

It remains to check that $\lambda$ satisfies the Hadamard condition, i.e. that 
\[
\Lambda= (\rho\circ E)^{*}\lambda (\rho\circ E)
\]
 satisfies $(\mu{\rm sc})$. By the well-known propagation  property of \cite{FSW}(see also \cite{SV}), it  suffices to check $(\mu{\rm sc})$ in $T^{*}\Omega\times T^{*}\Omega\backslash\{0\}$,  since $\Omega$
is a neighborhood of $S$ in $M$ by (\ref{Had.e1}).  Set: 
\[
\Lambda_{n}:= (\rho\circ E)^{*}\chi_{n}^{*}\lambda_{n}\chi_{n} (\rho\circ E),
\]
so that  $\Lambda= \sum_{n}\Lambda_{n}$. It suffices to check that $\Lambda_{n}$ 
satisfies $(\mu{\rm sc})$ in  $T^{*}\Omega\times T^{*}\Omega\backslash\{0\}$.

Using the support properties of $E$ and condition (\ref{Had.e1}) {\it (iii)}, we obtain that the restriction to $\Omega\times \Omega$ of the distribution kernel of $\Lambda_{n}$ is supported in $\tilde{\Omega}_{n}\times \tilde{\Omega}_{n}$. Therefore, up to diffeomorphisms, $\Lambda_n$ is equal to
\[
C_{n}:= (\rho\circ E_n)^{*}\chi_{n}^{*}\lambda_{n}\chi_{n} (\rho\circ E_n)
\]
on $\tilde{\Omega}_{n}\times \tilde{\Omega}_{n}$, where $E_n$ is the propagator associated to $P_n$. Using the invariance of the wavefront set under diffeomorphisms, it follows that $\Lambda_n$ satisfies the microlocal spectrum condition.  Therefore $\Lambda$ is the two-point function of a gauge-invariant quasi-free Hadamard state.

\subsubsection{Comparison between the two constructions}
 The FNW construction has the important advantage to generate pure Hadamard states for an arbitrary space-time from pure Hadamard states for an ultra-static one.  On the other hand, the relation (\ref{ascoli}) between the two two-point functions expressed in terms of Cauchy data is difficult to control in practice, since it involves the propagator $U'(-T, 0)$ for the intermediate metric $g'$, which is a non-local operator.

In contrast, our construction does  in general not generate a pure state, even if all the local states $\omega_{n}$ are so.  However the two-point function  given by (\ref{ascola}) is easier to control in practice, since it depends only on the two-point functions $\lambda_{n}$ and on the local operators $\chi_{n}$.
\appendix
\section{Various proofs}\label{sec-app1}

 \subsection{Proof of Lemma \ref{buda}.}\label{budaproof}
 
 Set $u(t,s):= \Texp({\textstyle\int_{s}^{t}}\i\epsilon(\sigma)d\sigma)$. We claim that it suffices to prove that 
 \begin{equation}
\label{e.buda1}
u(t,s)s_{-\infty}(t,s)\in \Psi^{-\infty}(\rr^{d}), \ \forall \  (t,s)\in \rr^{2}.
\end{equation}
In fact we have
 \[
\begin{array}{rl}
\p_{s}u(t,s)s_{-\infty}(t,s)=& u(t,s)\left(-\i \epsilon(s)s_{-\infty}(t,s)+\p_{s}s_{-\infty}(t,s)\right),\\[2mm]
\p_{t}u(t,s)s_{-\infty}(t,s)=& u(t,s)\left(\i u(s,t) \epsilon(t)u(t,s)s_{-\infty}(t,s)+\p_{t}s_{-\infty}(t,s)\right).
\end{array}
\]
We note that  $-\i \epsilon(s)s_{-\infty}(t,s)+ \p_{s}s_{-\infty}(t,s)\in \Psi^{-\infty}(\rr^{d})$, and by Prop. \ref{1.2} (2) $\i u(s,t)\epsilon(t)u(t,s)s_{-\infty}(t,s)+ \p_{t}s_{-\infty}(t,s)\in \Psi^{-\infty}(\rr^{d})$.  We can argue similarly to control higher derivatives in $(t,s)$.

We  first claim that 
\begin{equation}\label{e.buda0}
 \langle D_{x}\rangle^{m}u(t,s)\langle D_{x}\rangle^{-m}\in B(L^{2}(\rr^{d})), \ m\in \rr.
\end{equation}
In fact by Prop. \ref{1.2} we know that $u(s,t)\langle D_{x}\rangle^{m}u(t,s)\in \Psi^{m}(\rr^{d})$ and is moreover elliptic in this class, which proves (\ref{e.buda0}).

To prove (\ref{e.buda1}) we will use the Beals criterion  recalled in the proof of Prop. \ref{1.1}, and show that 
\begin{equation}
\label{e.buda3}
\langle D_{x}\rangle^{m}\ad_{x}^{\alpha}\ad_{D_{x}}^{\beta} (u(t,s)s_{-\infty}(t,s))\in B(L^{2}(\rr^{d})), \ \forall \ \alpha,\beta\in \nn^{d}, \ m\in \nn.
\end{equation}
We note that for $i= 1,\dots, d$:
\[
\begin{array}{rl}
[D_{i}, u(t,s)s_{-\infty}]=& u(t,s)(u(s,t)D_{i}u(t,s)- D_{i})s_{-\infty}+ u(t,s)[D_{i}, s_{-\infty}],\\[2mm]
[x_{i}, u(t,s)s_{-\infty}]=& u(t,s)(u(s,t)x_{i}u(t,s)- x_{i})s_{-\infty}+ u(t,s)[x_{i}, s_{-\infty}].
\end{array}
\]
By Prop. \ref{1.2} we know that $u(s,t)D_{i}u(t,s)- D_{i}\in \Psi^{1}(\rr^{d})$. On the other hand we have:
\[
\begin{array}{rl}
u(s,t)x_{i}u(t,s)- x_{i}=& \i \int_{s}^{t}u(s, \sigma)[\epsilon(\sigma), x_{i}]u(\sigma, s)d \sigma\\[2mm]
=& \int_{s}^{t} u(s, \sigma) a_{i}(\sigma) u(\sigma, s)d \sigma,
\end{array}
\]
where $a_{i}(\sigma)\in C^{\infty}(\rr, \Psi^{0}(\rr^{d}))$. Therefore we obtain that
\[
\begin{array}{l}
\ad_{D_{i}}u(t,s)s_{-\infty}=u(t,s)s_{-\infty, i}, \\[2mm]
 \ad_{x_{i}}u(t,s)s_{-\infty}= u(t,s)r_{-\infty, i}, \\[2mm]
  s_{-\infty, i}, \ r_{-\infty, i}\in\Psi^{-\infty}(\rr^{d}).
\end{array}
\]
Using also (\ref{e.buda0}), this implies (\ref{e.buda3}) by induction. \qed

\subsection{Proof of Lemma \ref{2.2}}\label{sec-app1.1bis}
Set $a(t)= a(t, \rx, D_{\rx})$. Since $a(t)$ is a second order differential operator,we have
\beq\label{e.bes1}
a(t)=a_{2}(t)+ a_{1}(t),\ a_{i}(t)\in \cinf(\rr, \Psi^{i}_{\rm ph}), \ a_{i}(t)= a_{i}(t)^{*}, \ i=1,2 
\eeq
and $a_{2}(t)= \sum_{ij}D_{i}a^{ij}(t, \rx)D_{j}$. From (\ref{e0.3}) we obtain that $a_{2}(t)\geq c(t)D^{2}$. Therefore we can find 
$r_{-\infty,1}(t)= r_{-\infty, 1}(t, D_{\rx})\in \cinf(\rr, \Psi^{-\infty})$ such that
\begin{equation}
\label{e.bes20}
a_{2}(t)- r_{-\infty,1}(t)\geq c(t)(D^{2}+\one).
\end{equation}
The operator $a_{2}(t)- r_{-\infty,1}(t)$ is elliptic in $\cinf(\rr, \Psi^{2}_{\rm ph})$ and strictly positive. By Prop. \ref{1.1},  $\epsilon_{1}(t):= (a_{2}(t)- r_{-\infty, 1}(t))^{\12}\in \cinf(\rr, \Psi^{1}_{\rm ph})$, and $\epsilon_{1}(t)$ is elliptic in $\Psi^{1}_{\rm ph}$ with principal symbol $(k_{i}a^{ij}(t, \rx)k_{j})^{\12}$. From (\ref{e.bes1}) we get
\begin{equation}
\label{e.bes2}
a(t)- r_{-\infty, 1}(t)= \epsilon_{1}^{2}(t)+ a_{1}(t)= \epsilon_{1}(t)(\one + s_{-1}(t))\epsilon_{1}(t),
\end{equation}
for $s_{-1}(t)= \epsilon_{1}(t)^{-1}a_{1}(t)\epsilon_{-1}(t)\in \cinf(\rr, \Psi^{1}_{\rm ph})$. We fix a cutoff function $\chi\in \cinf(\rr)$ with $\chi(s)\equiv 1$ for $|s|\geq 2$, $\chi(s)\equiv 0$ for $|s|\leq 1$.  Then
\[
\begin{array}{rl}
&\chi(R^{-1} |D_{\rx}|)s_{-1}(t)\chi(R^{-1} |D_{\rx}|)\in \cinf(\rr, \Psi^{1}_{\rm ph}),\\[2mm]
& s_{-1}(t)- \chi(R^{-1} |D_{\rx}|)s_{-1}(t)\chi(R^{-1} |D_{\rx}|)\in \cinf(\rr, \Psi^{-\infty}),\\[2mm]
&\lim_{R\to \infty}\chi(R^{-1} |D_{\rx}|)s_{-1}(t)\chi(R^{-1} |D_{\rx}|)=0\hbox{ in }B(L^{2}(\rr^{d})),
\end{array}
\]
where we used (\ref{e0.0}) in the last statement.
This implies that we can find $R= R(t)\gg 1$ such that:
\[
\begin{array}{rl}
&\chi(R |D_{\rx}|)s_{-1}(t)\chi(R |D_{\rx}|)=: \tilde{s}_{-1}(t)\in \cinf(\rr, \Psi^{-1}_{\rm ph}),\\[2mm]
&s_{-1}(t)- \tilde{s}_{-1}(t)=: \tilde{s}_{-\infty}(t)\in \cinf(\rr, \Psi^{-\infty}),\\[2mm]
&\one + \tilde{s}_{-1}(t)\geq (1-\delta)\one, \ 0<\delta<1.
\end{array}
\]
It follows that 
\[
a(t)- r_{-\infty, 1}(t)- \epsilon_{1}(t)\tilde{s}_{-\infty}(t)\epsilon_{1}(t)= \epsilon_{1}(t)(\one + \tilde{s}_{-1}(t))\epsilon_{1}(t)=: \tilde{a}(t),
\]
where $\tilde{a}(t)\in \cinf(\rr, \Psi^{2}_{\rm ph})$, $\tilde{a}(t)$ is elliptic in $\Psi^{2}$ with principal symbol $k_{i}a^{ij}(t, \rx)k_{j}$ and 
strictly positive. We set
\[
\begin{array}{rl}
r_{-\infty}^{\w}(t, \rx, D_{\rx}):=& r_{-\infty, 1}(t)-  \epsilon_{1}(t)\tilde{s}_{-\infty}(t)\epsilon_{1}(t)\in \cinf(\rr, \Psi^{-\infty}),\\[2mm]
\epsilon^{\w}(t, \rx, D_{\rx}):=& (\tilde{a}(t))^{\12}\in \cinf(\rr, \Psi^{1}_{\rm ph})
\end{array}
\]
Again by Prop. \ref{1.1} $\epsilon^{\w}(t, \rx, D_{\rx})$ has principal symbol $(k_{i}a^{ij}(t, \rx)k_{j})^{\12}$. This completes the construction of $\epsilon(t)$ and $r_{-\infty}(t)$.  The uniqueness modulo $\Psi^{-\infty}$ follows from the fact that $\epsilon^{\w}(t, \rx, D_{\rx})= a(t, \rx, D_{\rx})^{\12}$, hence the asymptotic expansion of its symbol is unique. \qed

\subsection{Proof of Thm. \ref{3.1}}\label{sec-app1.2}
We start by proving an auxiliary lemma.
\begin{lemma}\label{kilian}
 Let $F: \cinf(\rr, \Psi^{\infty}(\rr^{d}))\to \cinf(\rr, \Psi^{\infty}(\rr^{d}))$ a  map such that:
 \begin{equation}
\label{tata}
F: \cinf(\rr, \Psi^{0}_{({\rm ph})}(\rr^{d}))\to \cinf(\rr, \Psi^{-1}_{({\rm ph})}(\rr^{d})),
\end{equation}
 \beq\label{tirlititi}
b_{1}- b_{2}\in \cinf(\rr, \Psi^{-j}_{({\rm ph})}(\rr^{d}))\ \Rightarrow \ F(b_{1})- F(b_{2})\in \cinf(\rr, \Psi^{-j-1}_{({\rm ph})}(\rr^{d})), \ \forall \ j\in \nn.
\eeq
Let also $a\in \cinf(\rr, \Psi^{0}_{({\rm ph})}(\rr^{d}))$. Then there exists a solution $b\in \cinf(\rr, \Psi^{0}_{({\rm ph})}(\rr^{d}))$, unique modulo $\cinf(\rr, \Psi^{-\infty}(\rr^{d}))$ of the equation:
\beq\label{turlututu}
b= a+ F(b)\hbox{ mod }\cinf(\rr, \Psi^{-\infty}(\rr^{d})).
\eeq
\end{lemma}
\proof We first prove existence.  Set $b_{0}=a$, $b_{n}= a + F(b_{n-1})$, $n\geq 1$. Using (\ref{tirlititi}) we easily obtain by induction on $n$ that:
\begin{equation}
\label{e.vlad1}
b_{n}- b_{n-1}\in \cinf(\rr, \Psi^{-n}), \ n\geq 1.
\end{equation}

It follows that we can find $b\in \cinf(\rr, \Psi^{0})$ such that $b-b_{n}\in \cinf(\rr, \Psi^{-n})$, $\forall \ n\in \nn$. In fact it suffices to choose
\[
b\sim \sum_{n=0}^{\infty}(b_{n}- b_{n-1}).
\] 
Then
\[
b-a-F(b)= b-b_{n}+ F(b_{n-1})- F(b)\in \cinf(\rr, \Psi^{-n}),
\]
using (\ref{tirlititi}) and the fact that  $b-b_{n}\in \cinf(\rr, \Psi^{-n})$, $b-b_{n-1}\in \cinf(\rr, \Psi^{-n+1})$. 

Let us now prove uniqueness. If $b$, $\tilde{b}$ solve (\ref{turlututu}), then \[
b- \tilde{b}= F(b)- F(\tilde{b})\hbox{ mod }\cinf(\rr, \Psi^{-\infty}),
\]
hence $b- \tilde{b}\in \cinf(\rr, \Psi^{-1})$. By induction using (\ref{tirlititi}), we obtain that $b-\tilde{b}\in \cinf(\rr, \Psi^{-n})$, $\forall \ n \in \nn$.  The poly-homogeneous case is treated similarly.  \qed

\medskip

We now prove Thm. \ref{3.1}. The proof is divided in several steps.

{\it Step 1:}
 we first determine the operator $b(t)$, modulo $\cinf(\rr, \Psi^{-\infty})$. Set $u(t,s)=\Texp(\i \int_{s}^{t}b(\sigma)d\sigma)$, for $b(t)\in \cinf(\rr, \Psi^{1})$, $b(t)$ elliptic in $\Psi^{1}$ and $b(t)- b^{*}(t)\in \Psi^{0}$.
We have:
\[
\p_{t}u(t,s)= \i b(t)u(t,s), \ \p_{t}^{2}u(t,s)= - b^{2}(t)u(t,s)+ \i \p_{t}b(t)u(t,s).
\]
By Lemma \ref{2.2} we have
\[
(\p_{t}^{2}+ a(t)) u(t,s)= (\epsilon^{2}(t)- b^{2}(t)+ \i \p_{t}b(t)+ r_{-\infty}(t))u(t,s),
\]
with $r_{-\infty}(t)\in \cinf(\rr, \Psi^{-\infty})$.

Let us try to solve the equation
\begin{equation}
\label{etoto.1}
b^{2}- \epsilon^{2} = \i \p_{t}b\hbox{ mod }\cinf(\rr, \Psi^{-\infty}).
\end{equation}
We look for  a solution of (\ref{etoto.1}) of the form:
\[
b= \epsilon + b_{0}, \ b_{0}\in \cinf(\rr, \Psi^{0}).
\]
Since 
\[
b^{2}- \epsilon^{2}=   (\epsilon b_{0}+ b_{0}\epsilon)+ b_{0}^{2}=  (2 \epsilon b_{0}+[b_{0}, \epsilon])+ b_{0}^{2},
\]
we obtain that (\ref{etoto.1}) is equivalent to
\begin{equation}
\label{etoto.2}
\begin{array}{rl}
b_{0}=& (2\epsilon)^{-1} \i \p_{t}\epsilon+ 
(2\epsilon)^{-1}\left([\epsilon, b_{0}]+ \i\p_{t}b_{0}-b_{0}^{2}\right)\hbox{ mod }\cinf(\rr, \Psi^{-\infty})\\[2mm]
=:& (2\epsilon)^{-1}\i \p_{t}\epsilon+ F(b_{0})\hbox{ mod }\cinf(\rr, \Psi^{-\infty}).
\end{array}
\end{equation}
To solve (\ref{etoto.2}) we apply Lemma \ref{kilian}. Clearly $(2\epsilon)^{-1}\i \p_{t}\epsilon\in \cinf(\rr, \Psi^{0})$ and $F$ maps  $\cinf(\rr, \Psi^{0})$ into $\cinf(\rr, \Psi^{-1})$. Since
\[
\begin{array}{rl}
&F(b_{1})- F(b_{2})\\[2mm]
=&(2\epsilon)^{-1}\left([\epsilon, b_{1}- b_{2}]+ \i\p_{t}(b_{1}- b_{2})- (b_{1}^{2}- b_{2}^{2})\right)\\[2mm]
=&(2\epsilon)^{-1}\left([\epsilon, b_{1}- b_{2}]+ \i\p_{t}(b_{1}- b_{2})- (b_{1}- b_{2})b_{1}- b_{2}(b_{1}- b_{2})\right),
\end{array}
\]
we see that hypothesis (\ref{tirlititi}) also holds. Therefore we can find a solution to (\ref{etoto.1}) with:
\beq\label{etoto.2bis}
b(t)=  \epsilon(t)+ (2\epsilon)^{-1}\i \p_{t}\epsilon\hbox{ mod }\cinf(\rr, \Psi^{-1}).
\eeq

This proves condition {\it (i)} of the theorem. 

Note also that if $b$ a solution of (\ref{etoto.1}), then $-b^{*}$ also solves (\ref{etoto.1}), 
since $\epsilon= \epsilon^{*}$.
 Therefore  if 
 \[
 b_{+}(t)= b(t), \ b_{-}(t)= - b^{*}(t), \hbox{ and }u_{\pm}(t,s)= \Texp(\i\int_{s}^{t}b_{\pm}(\sigma)d\sigma),
 \] 
 we have
 \[
(\p_{t}^{2}+ a(t, \rx, D_{\rx}))u_{\pm}(t,s)= r_{-\infty, \pm}(t)u_{\pm}(t,s), \ r_{-\infty, \pm}(t)\in \cinf(\rr, \Psi^{-\infty}).
\]

{\it Step 2:}

we now solve, modulo smoothing errors, the Cauchy problem (\ref{e3.1}). For $f\in \cH'(\rr^{d})\otimes \cc^{2}$,
we look for approximate solutions of (\ref{e3.1}) of the form:
\beq\label{etoto.0}
u_{+}(t,s)\left(d_{+}(s)f_{0}+ n_{+}(s)f_{1}\right)+ u_{-}(t,s)\left(d_{-}(s)f_{0}+ n_{-}(s)f_{1} \right).
\eeq
We obtain the conditions:
\begin{equation}
\label{etoto.3}
\left\{\begin{array}{rl}
&d_{+}(s)+ d_{-}(s)=\one,\\
&b_{+}(s)d_{+}(s)+ b_{-}(s)d_{-}(s)=0,\\
&n_{+}(s)+ n_{-}(s)=0,\\
&b_{+}(s)n_{+}(s)+ b_{-}(s)n_{-}(s)=\one.
\end{array}
\right.
\end{equation}
We deduce from (\ref{etoto.2bis}) that $b_{\pm}$ are elliptic in $\Psi^{1}$, $b_{\pm}= \pm \epsilon + \Psi^{0}$ and $b_{\pm}^{(-1)}b_{\mp}=-\one + \Psi^{-1}$. Therefore the solutions of (\ref{etoto.3}) mod $\Psi^{-\infty}$ are given by:
\beq\label{etoto.5}
\left\{\begin{array}{rl}
d_{+}(s)=& (\one -b_{-}(s)^{(-1)}b_{+}(s))^{(-1)}, \\[2mm]
 d_{-}(s)=&(\one - b_{+}(s)^{(-1)}b_{-}(s))^{(-1)},\\[2mm]
 n_{+}(s)= &(b_{+}(s)- b_{-}(s))^{(-1)},\\[2mm]
 n_{-}(s)= &-n_{+}(s).
\end{array}\right.
\eeq
Note that it follows from (\ref{etoto.5}) that:
\begin{equation}
\label{etoto.00}
d_{+}(s)^{(-1)}n_{+}(s)= -b_{-}(s)^{(-1)}, \ d_{-}(s)^{(-1)}n_{-}(s)=-b_{+}(s)^{(-1)}\hbox{ mod }\Psi^{-\infty}.
\end{equation}
Therefore we can rewrite (\ref{etoto.0}) as 
\begin{equation}
\label{etoto.4}
U(t,s)f:= u_{+}(t,s)d_{+}(s)\left(f_{0}+ r_{+}(s)f_{1}\right)+ u_{-}(t,s)d_{-}(s)\left(f_{0}- r_{-}(s)f_{1}\right),
\end{equation}
for
\begin{equation}
\label{etoto.6}
r_{+}(s)= - b_{-}(s)^{(-1)}, \ r_{-}(s)= b_{+}(s)^{(-1)}\hbox{ mod }\Psi^{-\infty}.
\end{equation}
Since $b_{+}(s)= b(s)$, $b_{-}(s)= - b^{*}(s)$ if we choose:
\beq\label{etoto.6bis}
 r(s)= b^{*}(s)^{(-1)}\hbox{ mod }\Psi^{-\infty}, 
\eeq
and fix
\[
 r_{+}(s):= r(s), \ r_{-}(t):= r^{*}(s),
\]
then (\ref{etoto.6}) is satisfied. We now check that we can find $r(s)$ satisfying (\ref{etoto.6bis}) such that conditions {\it (iii)} and {\it (iv)} in the theorem are satisfied.

Let us denote $b(s)$, $r(s)$, $\epsilon(s)$ simply by $b$, $r$, $\epsilon$.  Since $b= \epsilon+ \Psi^{0}$ we have $r= \epsilon^{-1}+ \Psi^{-2}$, hence {\it (iii)} is satisfied. Moreover since 
 $\epsilon^{-\12}\in \Psi^{-\12}$ by Prop. \ref{1.1}, we have
\[
 r+ r^{*}= 2 \epsilon^{-1}+ \Psi^{-2}= \epsilon^{-\12}( 2\one + s_{-1}^{\rm w}(\rx, D_{\rx}))\epsilon^{-\12}, 
\]
where $s_{-1}(\rx, k)\in S^{-1}_{\rm ph}(\rr^{2d})$. We write
\[
\begin{array}{rl}
s_{-1}(\rx, k)= &s_{-1}(\rx, k)\chi(R^{-1}|k|\geq 1)+  s_{-1}(\rx, k)\chi(R^{-1}|k|\leq 1)\\[2mm]
=:& s_{-1, R}(\rx, k)+ s_{-\infty, R}(\rx, k).
\end{array}
\]
Note that $s_{-\infty, R}\in S^{-\infty}(\rr^{2d})$ and $s_{-1, R}$ tends to $0$ in $S^{0}(\rr^{2d})$ when $R\to +\infty$. By (\ref{e0.0}) it follows  that $2\one +s_{-1, R}^{\rm w}(\rx, D_{\rx})\sim \one$ for $R$ large enough. 
Therefore replacing $r$ by 
\[
\tilde{r}= r- \12 \epsilon^{-\12}s_{-\infty, R}^{\rm w}(\rx, D_{\rx})\epsilon^{-\12}= r+ \Psi^{-\infty},
\]
we can ensure {\it (iv)}, keeping (\ref{etoto.6bis}) satisfied. 

Collecting what we have done so far we have:
\[
\left\{
\begin{array}{rl}
&(\p_{t}^{2}+ a(t, \rx, D_{\rx}))U(t,s)f
= r_{-\infty, +}(t)u_{+}(t,s)d_{+}(s)(f_{0}+ r_{+}(s)f_{1})\\[2mm]
+& r_{-\infty, -}(t)u_{-}(t,s)d_{-}(s)(f_{0}- r_{-}(s)f_{1}),\\[2mm]
&U(s,s)f= f_{0}+ t_{-\infty, 0}(s)f,\\[2mm]
&\i^{-1}\p_{t}U(s,s)f=f_{1}+ t_{-\infty, 1}(s)f, 
\end{array}\right.
\]
where $r_{-\infty, \pm}$ and $t_{-\infty, i}$ belong to $\cinf(\rr, \Psi^{-\infty})$. Applying also  Lemma \ref{buda} to the operators  $r_{-\infty,\pm}(t)u_{\pm}(t,s)$, we obtain statement (1) of the theorem.

Finally  $\tilde{\phi}(t)= U(t,s)f$ solves
\[
\left\{
\begin{array}{rl}
&\p_{t}^{2}\tilde\phi(t)+a(t, \rx, D_{\rx})\tilde\phi(t)\in C^{\infty}(\rr, \cH(\rr^{d})),\\[2mm]
&\phi(s)-f_{0} \in \cH(\rr^{d}),\\[2mm]
&\i^{-1}\p_{t} \phi(s)-f_{1}\in \cH(\rr^{d}).
\end{array}
\right.
\]
By the uniqueness of the Cauchy problem (\ref{e3.1}) we obtain that $\phi(t)-\tilde{\phi}(t)\in C^{\infty}(\rr, \cH(\rr^{d}))$, which proves (2).

This completes the proof of the theorem. \qed

\end{document}